\documentclass[sigconf,nonacm]{acmart}

\usepackage[utf8]{inputenc} 
\usepackage[T1]{fontenc}    
\usepackage{url}            
\usepackage{booktabs}       
\usepackage{nicefrac}       
\usepackage{microtype}      
\usepackage{xcolor}         
\usepackage{graphicx}
\usepackage{caption}

\usepackage{amsmath}
\usepackage{amssymb}
\usepackage{subfig}
\usepackage{diagbox}
\usepackage{epsfig}
\usepackage{extpfeil}
\usepackage{wrapfig}
\usepackage{multirow}
\usepackage{wrapfig}
\usepackage{color,xcolor,colortbl}
\definecolor{lightgray}{gray}{0.95}
\definecolor{pink}{rgb}{1.0, 0.75, 0.8}
\definecolor{piggypink}{rgb}{0.99, 0.87, 0.9}
\definecolor{peachpuff}{rgb}{1.0, 0.85, 0.73}
\definecolor{color3}{gray}{0.95}
\definecolor{rouse}{rgb}{0.981,0.961,0.941}
\newcommand{\tsp}{^{\mathsf{T}}}

\usepackage{adjustbox}
\usepackage{booktabs}
\usepackage{multirow}
\usepackage{makecell}

\usepackage{array}

\definecolor{shadecolor}{rgb}{0.92,0.92,0.92}

\makeatletter %
\@namedef{ver@everyshi.sty}{}
\makeatother
\usepackage{tikz}

\newcommand{\name}{0}
\newcommand{\h}{0}
\newcommand{\w}{0.15}
\newcommand{\wa}{0.15}
\newlength \g

\AtBeginDocument{%
  \providecommand\BibTeX{{%
    \normalfont B\kern-0.5em{\scshape i\kern-0.25em b}\kern-0.8em\TeX}}}

\acmConference[Preprint]{2023}{2023}{Arxiv}
\begin{document}

\title{Degradation-Noise-Aware Deep Unfolding Transformer for Hyperspectral Image Denoising}

 \author{Haijin Zeng}
 \email{haijin.zeng@imec.be}

 \affiliation{%
   \institution{IMEC-UGent}
   \city{Ghent}
   \country{Belgium}
 }

  
 \author{Jiezhang Cao}
  \email{jiezhang.cao@vision.ee.ethz.ch}
   \affiliation{%
  	\institution{Computer Vision Lab, ETH}
  	\city{Zurich}
  	\country{Switzerland}
  }
  
 \author{Kai Feng}
  \email{fengkainpu@gmail.com}
   \affiliation{%
  	\institution{NPU}
  	\city{Xi'an}
  	\country{China}
  }
  
 \author{Shaoguang Huang}
   \affiliation{%
  	\institution{CUG}
  	\city{Wuhan}
  	\country{China}
  }
  
 \author{Hongyan Zhang}
   \affiliation{%
  	\institution{CUG}
\city{Wuhan}
\country{China}
}

 \author{Hiep Luong}
 \affiliation{%
	\institution{IMEC-UGent}
	\city{Ghent}
	\country{Belgium}
}
  
 \author{Wilfried Philips}
 \affiliation{%
	\institution{IMEC-UGent}
	\city{Ghent}
	\country{Belgium}
}

 \renewcommand{\shortauthors}{Trovato and Tobin, et al.}

\begin{abstract}
Hyperspectral imaging (HI) has emerged as a powerful tool in diverse fields such as medical diagnosis, industrial inspection, and agriculture, owing to its ability to detect subtle differences in physical properties through high spectral resolution. However, hyperspectral images (HSIs) are often quite noisy because of narrow band spectral filtering.
To reduce the noise in HSI data cubes, both model-driven and learning-based denoising algorithms have been proposed. 
However, model-based approaches rely on hand-crafted priors and hyperparameters, while learning-based methods are incapable of estimating the inherent degradation patterns and noise distributions in the imaging procedure, which could inform supervised learning. Secondly, learning-based algorithms predominantly rely on CNN and fail to capture long-range dependencies, resulting in limited interpretability.
This paper proposes a Degradation-Noise-Aware Unfolding Network (DNA-Net) that addresses these issues. Firstly, DNA-Net models sparse noise, Gaussian noise, and explicitly represent image prior using transformer. Then the model is unfolded into an end-to-end network, the hyperparameters within the model are estimated from the noisy HSI and degradation model and utilizes them to control each iteration. Additionally, we introduce a novel U-Shaped Local-Non-local-Spectral Transformer (U-LNSA) that captures spectral correlation, local contents, and non-local dependencies simultaneously. By integrating U-LNSA into DNA-Net, we present the first Transformer-based deep unfolding HSI denoising method.
Experimental results show that DNA-Net outperforms state-of-the-art methods, and the modeling of noise distributions helps in cases with heavy noise.

\end{abstract}


\begin{teaserfigure}
	\centering
	\renewcommand{\h}{0.105}
	\renewcommand{\wa}{0.12}
	\newcommand{\wb}{0.16}
	\renewcommand{\g}{-0.7mm}
	\renewcommand{\tabcolsep}{1.8pt}
	\renewcommand{\arraystretch}{1}
        \resizebox{1.0\linewidth}{!} {
		\begin{tabular}{cc}	
                \Huge
			\renewcommand{\name}{figures/TSA_S1to10/}
			\renewcommand{\h}{0.2}
			\renewcommand{\w}{0.2}
			\begin{tabular}{cc}
				\begin{adjustbox}{valign=t}
					\begin{tabular}{c}%
		         	\includegraphics[trim={0 0 0 0 },clip, width=0.3\textwidth]{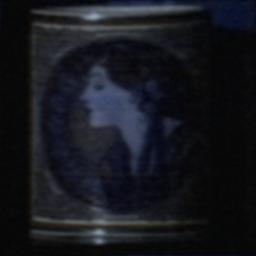}
						\\
						KAIST SCENE-02 \\
                        Gaussian Noise: $\mathcal{N}(0, 0.2)$ \\Sparse Noise: $p=0.1$
					\end{tabular}
				\end{adjustbox}
				\begin{adjustbox}{valign=t}
					\begin{tabular}{cccccccccc}
                            \includegraphics[trim={70 120 130 80  },clip,height=\h \textwidth, width=\w \textwidth]{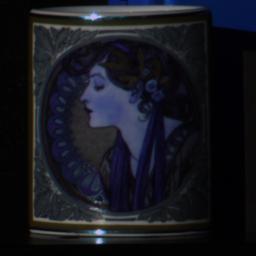} \hspace{\g} &
						\includegraphics[trim={70 120 130 80  },clip,height=\h \textwidth, width=\w \textwidth]{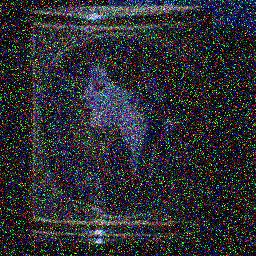} \hspace{\g} &
						\includegraphics[trim={70 120 130 80   },clip,height=\h \textwidth, width=\w \textwidth]{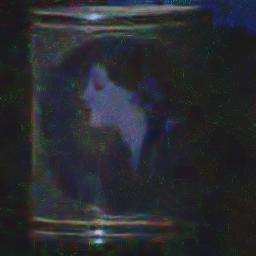} \hspace{\g} &
      					\includegraphics[trim={70 120 130 80  },clip,height=\h \textwidth, width=\w \textwidth]{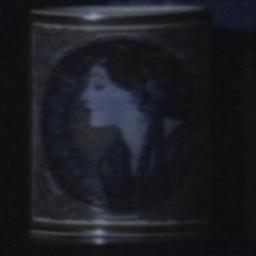} \hspace{\g} &
						\includegraphics[trim={70 120 130 80   },clip,height=\h \textwidth, width=\w \textwidth]{\name 3DTNN_FW_TSA_S2.jpg} \hspace{\g} &
						\includegraphics[trim={70 120 130 80   },clip,height=\h \textwidth, width=\w \textwidth]{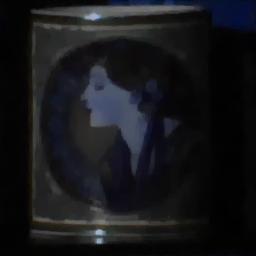} \hspace{\g} &
						\includegraphics[trim={70 120 130 80   },clip,height=\h \textwidth, width=\w \textwidth]{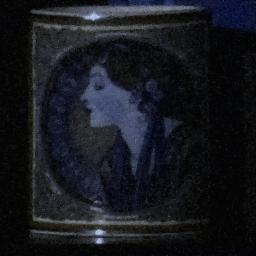} \hspace{\g} &
						\includegraphics[trim={70 120 130 80   },clip,height=\h \textwidth, width=\w \textwidth]{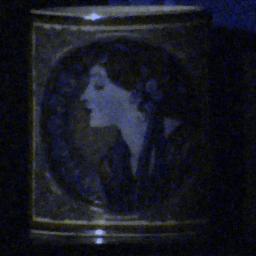} \hspace{\g} &
						\includegraphics[trim={70 120 130 80   },clip,height=\h \textwidth, width=\w \textwidth]{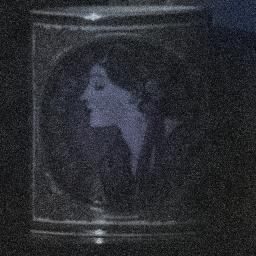} \hspace{\g} &
						\includegraphics[trim={70 120 130 80   },clip,height=\h \textwidth, width=\w \textwidth]{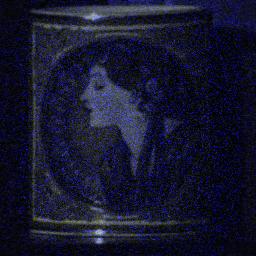} \hspace{\g} 
						\\
						GT \hspace{\g} &
						Noisy Image \hspace{\g} & BM3D~\cite{BM3D} \hspace{\g} & 3DTNN~\cite{3DTNN} \hspace{\g} & 3DTNN-FW~\cite{3DTNN_FW} & E3DTV~\cite{E3DTV} & FGSLR~\cite{FGSLR} & LRTV~\cite{LRTV} & NAILRMA~\cite{NAILRMA} & TLRLSSTV~\cite{TLR_LSSTV}
						\\
						\vspace{-2.5mm}
						\\
						\includegraphics[trim={70 120 130 80   },clip,height=\h \textwidth, width=\w \textwidth]{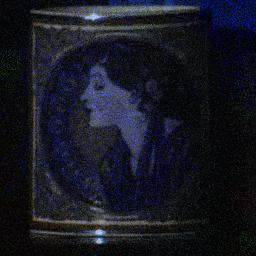} \hspace{\g} &
						\includegraphics[trim={70 120 130 80  },clip,height=\h \textwidth, width=\w \textwidth]{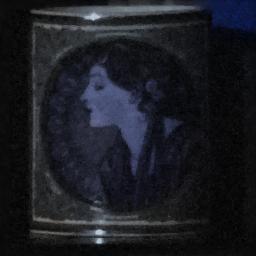} \hspace{\g} &
      					\includegraphics[trim={70 120 130 80   },clip,height=\h \textwidth, width=\w \textwidth]{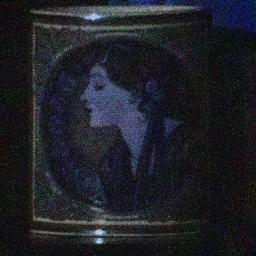} \hspace{\g} &
						\includegraphics[trim={70 120 130 80  },clip,height=\h \textwidth, width=\w \textwidth]{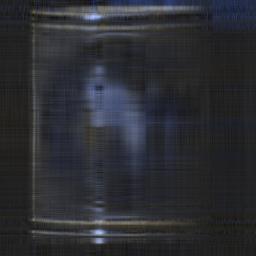} \hspace{\g} &
						\includegraphics[trim={70 120 130 80   },clip,height=\h \textwidth, width=\w \textwidth]{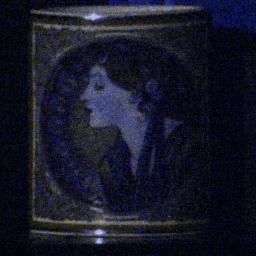}
						\hspace{\g} &		
						\includegraphics[trim={70 120 130 80   },clip,height=\h \textwidth, width=\w \textwidth]{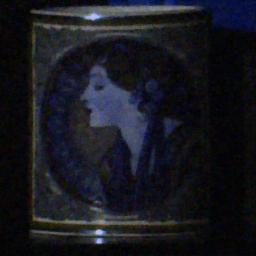} \hspace{\g} &		
						\includegraphics[trim={70 120 130 80   },clip,height=\h \textwidth, width=\w \textwidth]{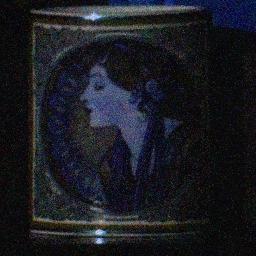} \hspace{\g} &
                            \includegraphics[trim={70 120 130 80   },clip,height=\h \textwidth, width=\w \textwidth]{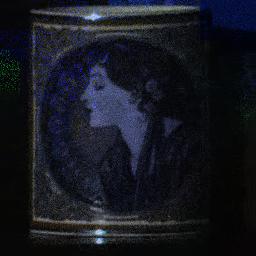} \hspace{\g} &
						\includegraphics[trim={70 120 130 80   },clip,height=\h \textwidth, width=\w \textwidth]{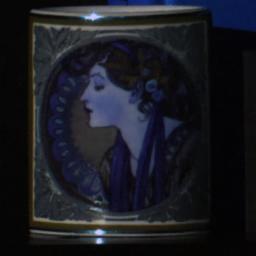}
						\hspace{\g} &
						\includegraphics[trim={70 120 130 80   },clip,height=\h \textwidth, width=\w \textwidth]{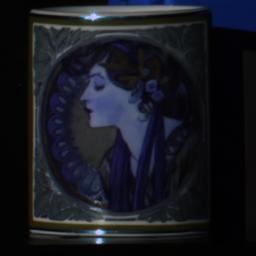} \hspace{\g} 
                            \\						
						LLRSSTV~\cite{LLRPnP} \hspace{\g} & LLxRGTV~\cite{LLxRGTV} \hspace{\g} & LRMR~\cite{LRMR} \hspace{\g} & LRTA~\cite{LRTA} & LRTDCTV~\cite{LRTDCTV} & LRTDTV~\cite{wang2017hyperspectral} & SSTV~\cite{SSTV} & LLRPnP~\cite{LLRPnP} \hspace{\g} & SERT~[cvpr2023]\cite{SERT} & \textbf{DNA-Net (Our)}
						\\
					\end{tabular}
				\end{adjustbox}
			\end{tabular}	
		\end{tabular}
	}
  \label{fig:teaser}
\end{teaserfigure}

\maketitle

\section{Introduction}
Hyperspectral image (HSI) comprises numerous bands that cover a broad spectrum range. Due to its ability to incorporate wavelengths beyond the visible spectrum and distinguish subtle differences in various materials, HSI offers significant advantages for various applications, including remote sensing, face recognition, medical diagnosis, and classification \cite{hsi_a1,hsi_a2,hsi_a3,hsi_a4}. However, current hyperspectral imaging techniques always produce HSIs with substantial degradation, e.g., electronic noise, quantization noise \cite{9732909,zhuang2018fast,SERT}, which emphasizes the importance of robust denoising algorithms to improve the image quality of HSI.

Research on HSI denoising can be categorized into four major directions: model-based models, Plug-and-play (PnP) algorithms, End-to-end (E2E) deep neural networks, and Deep unfolding methods. 
\textbf{Model-based methods} for HSI reconstruction rely on hand-crafted image priors such as sparsity based total variation, and spectral correlation based low-rank representation, among others. These methods have theoretical foundations and can be explained. However, they require manual parameter adjustment and also need lots of iterations to be converged, which slows down the reconstruction process. 
\textbf{PnP algorithms} employ pre-trained denoising network as Plug-and-Play priors of traditional model-based methods. However, the performance of PnP methods is limited by the fixed nature of the pre-trained networks, which cannot be re-trained and the network is not customised for the task.

\textbf{E2E approaches} for HSI denoising usually use a convolutional neural network (CNN), to learn the mapping function from a noisy HSI to the clean HSI. E2E methods have the advantages of deep learning. Nevertheless, they disregard the degradation mechanism and noise distributions of HSI by learning a brute-force mapping from the noisy hyperspectral images to the underlying noise free HSI. Additionally, E2E methods lack theoretical foundations, interpretability. 

\textbf{Deep unfolding methods} involve the utilization of a sequence of stages to map noisy HSI observations to clean HSI cubes. Typically, each stage incorporates a fidelity phase, followed by a single-stage network that learns the underlying denoiser prior \cite{SMDS-Net,xiong2021mac}. It presents a highly interpretable network architecture that explicitly characterizes both the image priors and degradation model. These methods offer a hybrid approach that combines the strengths of both deep learning and model-based techniques, leading to promising denoising performance. Nonetheless, 
current deep unfolding algorithms encounter two primary issues. Firstly, the iterative learning process in these methods is highly dependent on the degradation pattern of the hyperspectral image but current methods do not estimate the HSI degradation patterns and noise characteristics to adjust the denoising network in each iteration. Secondly, the existing deep unfolding methods predominantly employ CNNs, which exhibit limitations in capturing non-local self-similarity and long-range dependencies, both critical for HSI denoising.

Recently, the attention mechanism-based Transformer~\cite{vaswani2017attention} model has emerged as a promising solution for overcoming the drawbacks of CNNs. The Transformer's strong capability to model the interactions of global contexts or non-local spatial regions has resulted in its wide application in image processing. 
For instance, image
classification~\cite{liu2021swin,arnab2021vivit}, object detection~\cite{de_detr,to_1}, semantic segmentation~\cite{cao2021swin,ts_2}, human pose estimation~\cite{tokenpose,transpose}, image restoration~\cite{ipt,rformer}, \emph{etc.} However, the use of Transformer in HSI denoising is faced with two primary issues. Firstly, the computational complexity of global Transformer~\cite{global_msa} increases quadratically with spatial dimensions, making it unaffordable in some cases. Secondly, the receptive fields of local Transformer~\cite{liu2021swin} are limited to position-specific windows, which hinders the matching of highly related content tokens during self-attention computation \cite{DAUF}, and it also does not take into account the spectral correlation of HSI, 

In this paper, we propose a novel approach to address the problem of HSI denoising by designing a Degradation-Noise-Aware Unfolding Network (DNA-Net) based on the maximum a posteriori (MAP) theory. Unlike previous deep unfolding methods, DNA-Net accurately models sparse noise and Gaussian noise using $l_1$-norm and F-norm, respectively. Moreover, it adaptively estimates informative hyperparameters from degraded noisy observations. These hyperparameters are then fed into each iteration of DNA-Net to adaptively scale the noise reduction.

Additionally, a new U-shaped local-non-local-spectral Transformer (U-LNSA) is proposed as the sub-denoiser in each iteration. U-LNSA jointly extracts local contextual information, models non-local dependencies and spectral correlation of HSI, while being computationally efficient. To achieve this, a spatial-spectral-split Multi-head Self-Attention block is customized to compose the basic unit of U-LNSA. Specifically, LNSA has three parallel attention branches, which calculate the self-attention within the local window, capture cross-window interactions by shuffling the tokens, and model the spectral information, respectively.

Finally, U-LNSA is integrated into DNA-Net to form an iterative architecture called the Degradation-Noise-Aware Unfolding Transformer. Experiments demonstrate that DNA-Net outperforms latest state-of-the-art (SOTA) methods by over 1.49dB, as illustrated in the teaser. 
Our contributions are summarized as follows:
\begin{enumerate}
    \item We propose an MAP-based unfolding network for HSI denoising that accurately models the degradation pattern and noise distributions of HSI, while incorporating adaptive hyperparameter estimation.

    \item We propose a novel Transformer U-LNSA and plug it into DNA-Net to do HSI denoising. To the best of our knowledge, DNA-Net is the first Transformer-based deep unfolding method for HSI denoising. 

    \item DNA-Net achieves significant performance improvements over state-of-the-art methods, particularly in scenarios involving heavy noise.
\end{enumerate}

\vspace{-3mm}
\section{Related Work}
\vspace{-1mm}

In recent years, a substantial body of research has been devoted to mitigating the challenge of denoising hyperspectral images. The literature on this topic can be broadly classified into two principal categories: conventional, which encompasses optimization-based models, and deep-learning-based approaches.

\subsection{Optimization and PnP-based Methods}

Optimization-based methods have been extensively employed for HSI denoising, with full-rank, low-rank matrix, and low-rank tensor approaches being the most commonly used techniques. Full-rank approaches utilize wavelet-based methods, combined with spatial and/or spectral handcrafted regularizers, to effectively capture the spatial and spectral dependencies of the HSI. On the other hand, low-rank approaches rely on low-rank constraints, such as the nuclear norm, to exploit the high spectral correlation or local smoothness present in the HSI. To sum up, traditional optimization-based methods formulate the problem as an optimization objective that takes into account the unique properties of the spectrum and images. Various handcrafted regularization terms, such as total variation, wavelet constraint, and low-rank representation  \cite{yuan2012hyperspectral,wang2017hyperspectral,othman2006noise,zhao2020fast,sun2017hyperspectral,wei2019low,maggioni2012nonlocal,peng2014decomposable}, have been incorporated in the optimization framework to enhance the denoising performance. These optimization-based methods are flexible enough to remove different types of noise \cite{he2015total,chen2017denoising} and can be extended to tasks beyond denoising \cite{chang2020weighted,fu2017adaptive,he2019non,pengE3DTV}. However, the performance of these methods is limited by the degree of matching between the handcrafted regularization and the underlying properties of the HSI, especially for complex real scenes \cite{he2015total,chen2017denoising}. To address this issue, plug-and-play methods integrate optimization-based methods with a learning-based prior \cite{DPLRTA,chan2016plug,danielyan2010image,zoran2011learning}, such as a plug-and-play Gaussian denoiser, to handle complex noise types that cannot be easily modeled by handcrafted regularization \cite{dabov2007image,zhang2018ffdnet}. These methods can remove different types of noise and can also be extended to tasks beyond denoising, e.g., \cite{liu2021hyperspectral,ma2020hyperspectral,ma2020hyperspectral,zhang2018ffdnet}.

\vspace{-2mm}
\subsection{ CNN and Unfolding based Methods}

Recently, several deep architectures have been proposed that take advantage of advances in deep learning for HSI denoising 
\cite{zhang2017beyond,chang2018hsi,yuan2018hyperspectral,dong2019deep,wei20203,lai2022deep,bodrito2021trainable,pang2022trq3dnet,wang2022uformer,chen2022hider,fu2022nonlocal,rui2022hyper,smds_net,lai2023mixed}. 
In \cite{yuan2018hyperspectral}, a spatio-spectral deep residual CNN was proposed that utilizes 3D and 2D convolutional filters to capture the dependencies of the images. The MemNet \cite{tai2017memnet} network and one variation MemNetRes \cite{coquelin2022hyde} (which is a combination of MemNet with the Hyres approach [61]), can provide competitive hyperspectral denoising results.
Inspired by the success of the 2D image denoising network DnCNN \cite{zhang2017beyond}, Chang \emph{et al.} \cite{chang2018hsi} proposed HSI-DeNet, which learns multi-channel 2-D filters to model spectral correlation. Yuan \emph{et al.} \cite{yuan2018hyperspectral} introduced a residual network structure with a sliding window strategy for remote sensed HSI. To further exploit spatial-spectral correlation, Dong \emph{et al.} \cite{dong2019deep} designed a 3D U-net architecture.
Although these methods have been successfully applied to various HSIs, most of them are limited to exploring inter-spectral correlations, which are significantly important for HSI denoising. To address this issue, the 3D Quasi-Recurrent Neural Network (QRNN3D) \cite{wei20203} employs 3D convolution components and quasi-recurrent pooling functions to capture the spatio-spectral dependencies of the HSI. Li et al. \cite{SERT} introduces a novel approach for HSI denoising, termed as Spectral Enhanced Rectangle Transformer (SERT). This method effectively leverages the non-local spatial similarity as well as the global spectral low-rank characteristic of noisy images, resulting in new SOTA denoising performance.

\vspace{-1mm}
\section{Proposed Method} \label{sec:method}
\subsection{Degradation Model of HSI}
\vspace{-1mm}
In HSI imaging system, we denote the observed 3D HSI and the latent noise free HSI as tensor $\mathcal{Y} \in \mathbb{R}^{M\times N \times P}$ and $\mathcal{X} \in \mathbb{R}^{M\times N \times P}$, respectively, where $P$ is the total number of spectral bands, $M$, $N$ denote the HSI's spatial height, width.  
Previous deep neural network based denoising approaches mainly focus on learning a deep blind model with paired training data, to reconstruct a hyperspectral image $\hat{\mathcal{X}}$ by referencing the observed noisy image $\mathcal{Y}$.
Mathematically, deep blind model is equivalent to maximize a posterior (MAP):
\begin{equation}
\vspace{-1mm}
    \min_{\theta} \mathcal{L} (\hat{\mathcal{X}}(\mathcal{Y}, \theta ), \mathcal{X} ), s.t., 
    \hat{\mathcal{X}}=\arg \min _{\mathcal{X}} F(\mathcal{X}, \mathcal{Y})+ \Phi (\mathcal{X}),
\end{equation}
where $F(\mathcal{X}, \mathcal{Y})$ denotes image fidelity term, $\Phi(\mathcal{X})$ is the regularization used to represent the prior of HSI, $\mathcal{L}(\cdot)$ denotes the loss function, $\theta$ is the learn-able parameters of network.
However, they do not estimate the HSI degradation patterns or model noise characteristics to guide the learning phase.

HSI are typically degraded by various types of noise, including Gaussian noise, sparse noise (such as impulse noise and stripes), and Poissonian noise \cite{zhuang2018fast,LRTV}. Meanwhile, Poissonian noise has emerged as a major concern in real HSI due to the decrease in spectral bandwidth resulting from an increase in the number of spectral bands in new-generation hyperspectral sensors \cite{hsi_noise_1,hsi_noise_2,hsi_noise_3}. This reduction in spectral bandwidth results in each spectral channel capturing fewer photons, leading to higher levels of Poissonian noise. However, if the mean value of the photon counts is greater than four, Poissonian noise can be approximated as additive Gaussian noise with nearly constant variance by using variance-stabilizing transformations \cite{hsi_noise_4}. Therefore, in this paper, we assume that the observation noise is additive Gaussian and sparse noise, then formulate the degradation model of HSI as follows:
\vspace{-2mm}
\begin{equation}
	\mathcal{Y} = \mathcal{X}+\mathcal{N} +\mathcal{S},
 \label{equ:degradation_model}
\end{equation}
where, $\mathcal{N} \in \mathbb{R}^{M\times N \times P}$ and $\mathcal{S} \in \mathbb{R}^{M\times N \times P}$ represent the imaging Gaussian noise and the sparse noise on the observation, respectively. Then the task of HSI denoising is given noisy HSI $\mathcal{Y}$ to reconstruct the noisy free HSI $\mathcal{X}$.

\begin{figure*}[t]
	\begin{center}
		\begin{tabular}[t]{c} \hspace{-2.4mm}
			\includegraphics[width=0.98\textwidth]{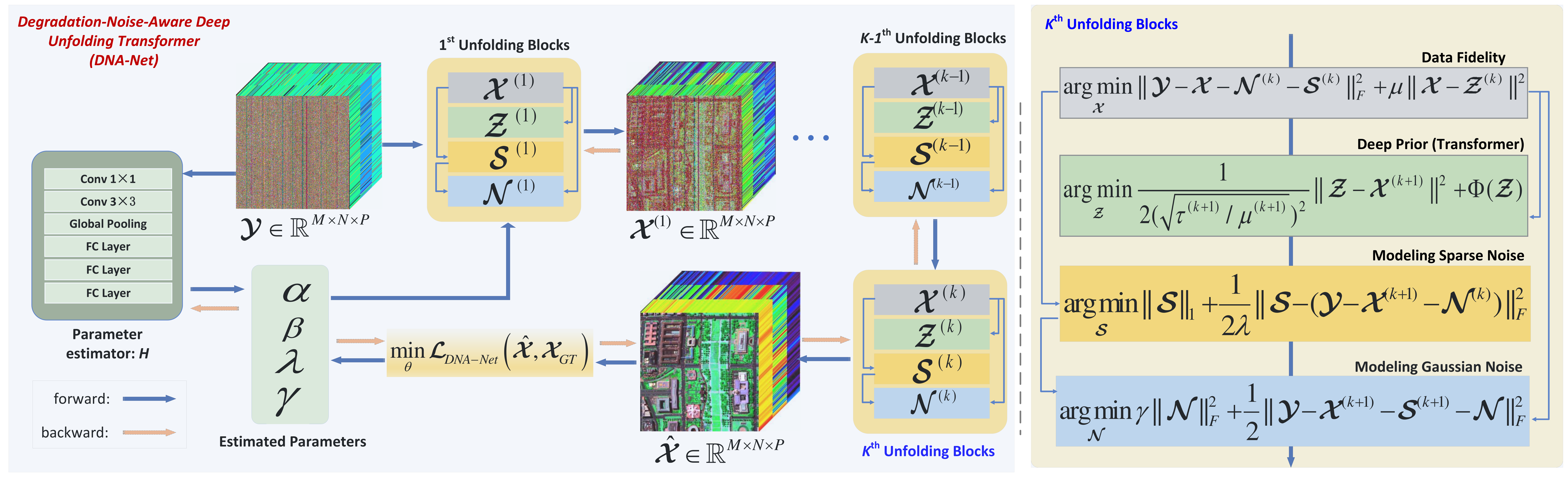}
		\end{tabular}
	\end{center}
	\vspace*{-6mm}
	\caption{\small The architecture of our DNA-Net with $K$ stages (iterations). $\textbf{H}$ estimates informative parameters from the noisy HSI $\mathcal{Y}$ and degradation matrix $\mathbf{I}$. The estimated parameters $\boldsymbol{\alpha}$, $\boldsymbol{\beta}$, $\boldsymbol{\lambda}$, and $\boldsymbol{\gamma}$ are fed into each stage of subsequent iterative learning.}
	\label{fig:pipeline}
	\vspace{-3mm}
\end{figure*}

\vspace{-1mm}
\subsection{Degradation-Noise-Aware Unfolding Network} \label{sec:dauf}
Based on degradation (noise) model (\ref{equ:degradation_model}), we then formulate a principled Degradation Noise Aware Unfolding Network (DNA-Net) as depicted in Fig.~\ref{fig:pipeline}. DNA-Net starts from the MAP theory. In particular, the original HSI signal could be estimated by minimizing the following energy function as follows:
\begin{equation}
\text{arg}~\underset{\mathcal{X},\mathcal{N},\mathcal{S}}{\text{min}}~~\frac{1}{2} || \mathcal{Y} - \mathcal{X} - \mathcal{N} - \mathcal{S} ||_{\text{F}}^2 + \tau \Phi(\mathcal{X}) + \lambda \|\mathcal{S}\|_{1}+\gamma \|\mathcal{N}\|_{\text{F}}^{2}, 
\label{eq:energy_1}
\end{equation}
where $\frac{1}{2} || \mathcal{Y} - \mathcal{X} - \mathcal{N} - \mathcal{S} ||^2$ is the data  fidelity term, $\Phi(\mathcal{X})$ is the image prior term, $\|\mathcal{S}\|_{1}, \|\mathcal{N}\|_{\text{F}}^{2}$ are the modeling of Gaussian noise and Sparse noise, and $\tau, \lambda, \gamma$ are hyper-parameters balancing the importance. By introducing an auxiliary variable $\mathcal{Z}$, the sub-problem of $\mathcal{X}$ in Eq.~\eqref{eq:energy_1} can be reformulated as 
\begin{equation}
\begin{aligned}
    \text{arg}~\underset{\mathcal{X},\mathcal{Z}}{\text{min}}~~ & \frac{1}{2} || \mathcal{Y} - \mathcal{X} - \mathcal{N} - \mathcal{S} ||_{\text{F}}^2 + \tau \Phi(\mathcal{Z}),
    ~ s.t.~\mathcal{Z} = \mathcal{X}.
\end{aligned}
\label{eq:energy_2}
\end{equation}
This is a constrained optimization problem, it can be solved directly, but to obtain an unfolding inference, we adopt half-quadratic splitting (HQS) algorithm for its simplicity and fast convergence. Then Eq.~\eqref{eq:energy_2} is solved by minimizing 
\begin{equation}
\begin{aligned}
    \mathcal{L}_{\mu,\lambda,\beta}(\mathcal{X},\mathcal{N},\mathcal{S}) = & \frac{1}{2} || \mathcal{Y} - \mathcal{X} - \mathcal{N} - \mathcal{S} ||_{\text{F}}^2 + \tau \Phi(\mathcal{Z}) + \lambda \|\mathcal{S}\|_{1} \\
    &+\gamma \|\mathcal{N}\|_{\text{F}}^{2} + \frac{\mu}{2} || \mathcal{Z} - \mathcal{X} ||^2,
\end{aligned}
\label{eq:hqs_1}
\end{equation}
where $\mu$ is a penalty parameter that forces $\mathcal{X}$ and $\mathcal{Z}$ to approach the same fixed point. Subsequently, Eq.~\eqref{eq:hqs_1} can be solved by decoupling $\mathcal{X}$ and $\mathcal{Z}$  into the following two iterative sub-problems as
\begin{equation}
\begin{aligned}
	&\mathcal{X}^{(k+1)} = \text{arg}~\underset{\mathcal{X}}{\text{min}}~~|| \mathcal{Y} - \mathcal{X} - \mathcal{N}^{(k)} - \mathcal{S}^{(k)} ||_{\text{F}}^2 + \mu || \mathcal{X} - \mathcal{Z}^{(k)} ||^2~,\\
 & \mathcal{Z}^{(k+1)} = \text{arg}~\underset{\mathcal{Z}}{\text{min}}~~\frac{\mu}{2} || \mathcal{Z} - \mathcal{X}^{(k+1)} ||^2 +  \tau \Phi(\mathcal{Z}),
 \end{aligned}
\label{eq:hqs_2}
\vspace{-2mm}
\end{equation}
where $k = 0, 1,~...~, K-1$ indexes the iteration. 

\noindent 1) Update $\mathcal{X}$. For the data fidelity term, \emph{i.e.}, $\mathcal{X}_{k+1}$ in Eq.~\eqref{eq:hqs_2}, it is associated with a quadratic regularized least-squares problem, which has a closed-form solution, 
\begin{equation} \label{equ:solution_x}
    \mathcal{X}^{(k+1)}=\frac{\mathcal{Y}-\mathcal{N}^{(k)}-\mathcal{S}^{(k)}
+\mu \mathcal{Z}^{(k)}}{1+\mu}.
\end{equation}

\noindent 2) Update $\mathcal{Z}$.
Extracting all terms containing $\mathcal{Z}$ from the augmented Lagrangian function, one can get:
\begin{equation}
\begin{aligned}
&\mathcal{Z}^{(k+1)}= \arg\min_{\mathcal{Z}} \frac{\mu}{2\tau}\left\|\mathcal{Z} - \mathcal{X}^{(k+1)}\right\|_{\text{F}}^{2} + \Phi(\mathcal{Z}).\\
\end{aligned}
\end{equation}
We also set $\tau$ as iteration-specific parameters and $\mathcal{Z}_{k+1}$ can be reformulated as 
\begin{equation}
        \label{eq:10}
	\mathcal{Z}^{(k+1)} = \text{arg}~\underset{\mathcal{Z}}{\text{min}}~~\frac{1}{2(\sqrt{\tau^{(k+1)}/\mu^{(k+1)}})^2}~|| \mathcal{Z} -\mathcal{X}^{(k+1)} ||^2 +  \Phi(\mathcal{Z}).
\end{equation}
From the perspective of Bayesian probability, Eq.~\eqref{eq:10} is equivalent to denoising image $\mathcal{X}^{(k+1)}$ with an Gaussian noise at level $\sqrt{\tau^{(k+1)}/\mu^{(k+1)}}$~\cite{pnp_3}. Here, we customised an U-shaped transformer network (see Sec. \ref{sec.ULNSA} for details) to act as the denoiser, i.e.,
\begin{equation} \label{equ:solution_Z}
    \mathcal{Z}^{(k+1)} = \mathcal{D}(\mathcal{X}^{(k+1)}, \sqrt{\tau^{(k+1)}/\mu^{(k+1)}}).
\end{equation}
3)	Update $\mathcal{S}$.
By extracting all items related to $\mathcal{S}$ from equation (\ref{eq:hqs_1}), 
\begin{equation}
\begin{aligned}
& \mathcal{S}^{(k+1)}=  \arg\min_{\mathcal{S}} \|\mathcal{S}\|_{1}+\frac{1}{2 \lambda}\left\|\mathcal{S} -\left(\mathcal{Y}-\mathcal{X}^{(k+1)}-\mathcal{N}^{(k)}\right)\right\|_{\text{F}}^{2}.
\end{aligned}
\end{equation}
By introducing the so-called soft-thresholding operator:
\begin{equation}
\operatorname{R}_{\Delta}(\mathbf{x})=
\left\{
\begin{array}{cc}
{x-\Delta,} & {\text { if } \quad x>\Delta} \\
{x+\Delta,} & {\text { if } \quad x<\Delta} \\
{0,} & {\text { otherwise }}
\end{array}
\right.
\end{equation}
where $x \in \mathbb{R}$ and $\Delta>0$. Then we can update $\mathcal{S}^{(k+1)}$ as
\begin{equation} \label{equ:solution_s}
\mathcal{S}^{(k+1)}=\operatorname{R}_{\lambda}\left(\mathcal{Y}-\mathcal{X}^{(k+1)}-\mathcal{N}^{(k)}\right).
\end{equation}

\noindent 4) Update $\mathcal{N}$, according to equation (\ref{eq:hqs_1}), we have
\begin{equation} 
\vspace{-2mm}
 \mathcal{N}^{(k+1)}=\arg\min_{\mathcal{N}} \frac{1}{2} || \mathcal{Y} - \mathcal{X} - \mathcal{N} - \mathcal{S} ||^2 +\gamma \|\mathcal{N}\|_{\text{F}}^{2}, 
\vspace{-0.5mm}
\end{equation}
it has a closed-form solution, 
\vspace{-1mm}
\begin{equation} \label{equ:solution_n}
\mathcal{N}^{(k+1)}=\frac{\mathcal{Y}-\mathcal{X}^{(k+1)}-\mathcal{S}^{(k+1)}}{1+2 \gamma}.
\vspace{-1mm}
\end{equation}

\noindent \textbf{Parameters estimation}. The hyperparameters $\tau, \lambda, \gamma$ on Eq.~\eqref{equ:solution_x}, Eq.~\eqref{equ:solution_Z}, Eq.~\eqref{equ:solution_s}, and Eq.~\eqref{equ:solution_n}, are used to weight the regularization terms to adjust the noise reduction strength on the noisy input, in order to minimize negative effects caused by noisy elements. However, the success of this weighting scheme is contingent upon prior knowledge and assumptions regarding the distribution of HSI noise. Consequently, these assumptions may not generalize well to diverse practical scenarios and can be biased towards complex real noises. To overcome this issue, we use a new data-driven approach for generalizing the weighting principle. We introduce an explicit hyperweight estimation network that maps an input noisy image to appropriate weights for regularization, which consists of a $conv1\times1$, a strided $conv3\times3$, a global average pooling, and three fully connected layers, as shown in Fig. \ref{fig:pipeline}. This method seeks to capture the general weighting principle in a more objective manner. Specifically, based on Equation \eqref{eq:hqs_2}, the penalty parameter $\mu$ must satisfy a certain level of magnitude to ensure that the fixed points of $\mathcal{X}$ and $\mathcal{Z}$ converge closely. As such, $\mu$ serves as a crucial parameter that governs the convergence behavior and output quality of each iteration. To mitigate the need for manual tuning of $\mu$, we automatically estimates $\mu$ based on a series of iteration-specific parameters derived from the HSI data. Specifically, we denote the value of $\mu$ in the $k$-th iteration as $\mu_k$. Then, for estimating the parameters of Eq.~\eqref{equ:solution_Z}, we set $\frac{1}{(\sqrt{\tau^{(k+1)}/\mu^{(k+1)}})^2} = \mu^{(k+1)}/\tau^{(k+1)}$ as parameters to be estimated from HSI. Let $\alpha^{(k)} \stackrel{\rm def}{=} \mu^{(k)}$, $\boldsymbol{\alpha} \stackrel{\rm def}{=} [\alpha^{(1)}, ... ,\alpha^{(k)}]$, $\beta^{(k)} \stackrel{\rm def}{=} \mu^{(k)}/\tau^{(k)}$, and $\boldsymbol{\beta} \stackrel{\rm def}{=} [\beta^{(1)}, ... ,\beta^{(k)}]$, $\boldsymbol{\gamma} \stackrel{\rm def}{=} [\gamma^{(1)}, ... ,\gamma^{(k)}]$, $\boldsymbol{\lambda} \stackrel{\rm def}{=} [\lambda^{(1)}, ... ,\lambda^{(k)}]$. Then we can formulate our DNA-Net as an iterative scheme: 
\begin{equation}
\vspace{-1.5mm}
\begin{aligned}
    &(\boldsymbol{\alpha}, \boldsymbol{\beta}, \boldsymbol{\gamma}, \boldsymbol{\lambda}) =  \mathcal{H}(\mathcal{Y}, \mathbf{I}),
    ~~~~\mathcal{X}^{(k+1)} = \mathcal{F}(\mathcal{Y}, \mathcal{Z}^{(k)}, \alpha^{(k+1)}, \mathbf{I}),\\
    &~~~~\mathcal{Z}^{(k+1)} = \mathcal{D}(\mathcal{X}^{(k+1)}, \beta^{(k+1)}), 
    ~~~~\mathcal{S}^{(k+1)} = \mathcal{R}(\mathcal{X}^{(k+1)}, \mathcal{N}^{(k)}, \lambda^{(k+1)}), \\
    &~~~~\mathcal{N}^{(k+1)} = \mathcal{G}(\mathcal{X}^{(k+1)}, \mathcal{S}^{(k+1)}, \gamma^{(k+1)}),
    \label{eq:para}
\end{aligned}
\end{equation}
where $\mathcal{H}$ denotes the parameter estimator that takes the noisy observation $\mathcal{Y}$ and the identity degradation matrix $\mathbf{I}$ of the HSI system as inputs, $\mathcal{F}, \mathcal{R}, \mathcal{G} $,  equivalent to Eq.~\eqref{equ:solution_x}, Eq.~\eqref{equ:solution_s}, Eq.~\eqref{equ:solution_n}, and $\mathcal{D}$ represents the Gaussian denoiser solving Eq.~\eqref{equ:solution_Z}.

\begin{figure*}[t]
	\begin{center}
		\begin{tabular}[t]{c} \hspace{-3.4mm}
			\includegraphics[width=0.99\textwidth]{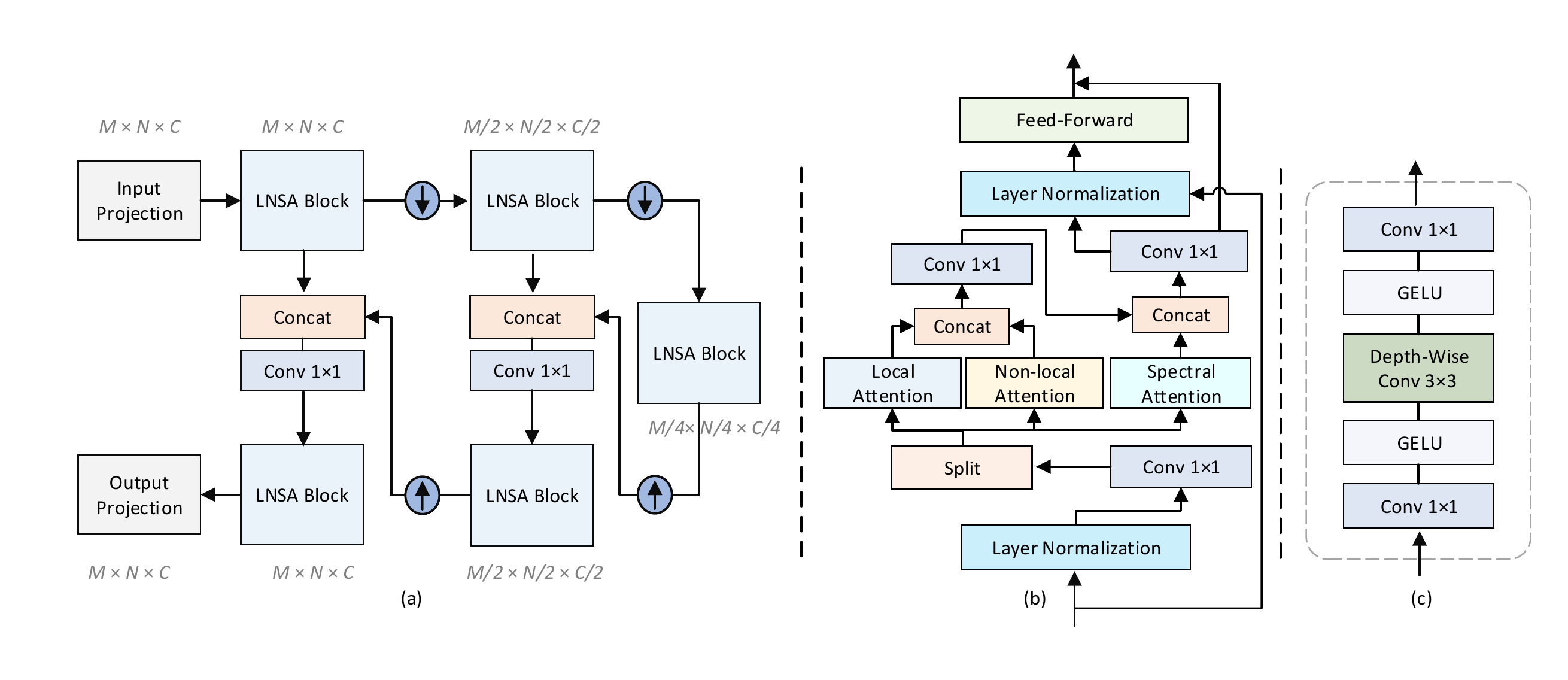}
		\end{tabular}
	\end{center}
	\vspace*{-12mm}
	\caption{\small Diagram of the proposed transformer denoiser U-LNSA for our unfolding network DNA-Net. (a) U-LNSA adopts a U-shaped structure, (b) LNSA block, (c) Feed-forward network.}
	\label{fig:U-LNSA}
	\vspace{-4mm}
\end{figure*}

\vspace{-2mm}
\subsection{Customised HSI Transformer Denoiser} \label{sec.ULNSA}
\vspace{-0.5mm}
Previous approaches for designing denoiser priors for RGB image restoration have mainly utilized convolutional neural networks (CNNs), which have shown limited capacity to capture long-range dependencies \cite{DAUF} when it comes to HSI denoising task. While local and global Transformers can model long-range dependencies effectively, directly applying them to hyperspectral image (HSI) denoising faces two challenges: limited receptive fields and high computational costs. Moreover, the high spectral resolution of HSI suggests strong spectral correlation among spectral modes, which differs significantly from classic RGB images. Therefore, it is also essential for the denoiser to model this critical feature of HSI.
To address these challenges, we propose a novel U-shaped Local-Non-local-Spectral Transformer (U-LNSA) denoiser that utilizes a Local-Non-local-Spectral Attention (LNSA) mechanism. This approach combines local and non-local attention mechanisms with spectral convolution to capture long-range dependencies and strong spectral correlations while maintaining reasonable computational costs. 

\noindent \textbf{Overview of Denoiser U-LNSA.}  
The architecture of U-LNSA is depicted in Fig.\ref{fig:U-LNSA} (a) and is built in a U-Net style by utilizing the proposed LNSA block as its fundamental unit. Given an input $\mathcal{X}^{(k+1)} \in \mathbb{R}^{M \times N \times C}$, U-LNSA first applies a $conv3\times3$ to map the concatenated reshaped $\mathcal{X}^{(k+1)}$ and stretched $\beta^{(k+1)}$ into feature $\mathcal{X}_0\in \mathbb{R}^{M\times N \times C}$. Next, $\mathcal{X}_0$ is fed through the encoder, bottleneck, and decoder to be embedded into the deep feature $\mathcal{X}_d\in \mathbb{R}^{M\times N\times C}$. Each level of the encoder or decoder includes an LNSA and a resizing module. Fig.\ref{fig:U-LNSA} (b) shows that LNSA comprises of two layer normalization (LN) operations, an LNSA, and a Feed-Forward Network (FFN) that is explained in Fig.~\ref{fig:U-LNSA} (c). The downsampling and upsampling modules are implemented as strided $conv4\times4$ and $deconv2\times2$, respectively. Finally, a $conv3\times3$ is applied to $\mathcal{X}_d$ to generate the residual image $\mathcal{T}^{(k+1)}\in \mathbb{R}^{M\times N\times C}$, and the output denoised image $\mathcal{Z}^{(k+1)}$ is obtained by adding reshaped $\mathcal{T}^{(k+1)}$ to $\mathcal{X}^{(k+1)}$: $\mathcal{Z}^{(k+1)}=\mathcal{T}^{(k+1)}+\mathcal{X}^{(k+1)}$.

\begin{figure}[t]
        \vspace{-3mm}
			\includegraphics[width=0.5\textwidth]{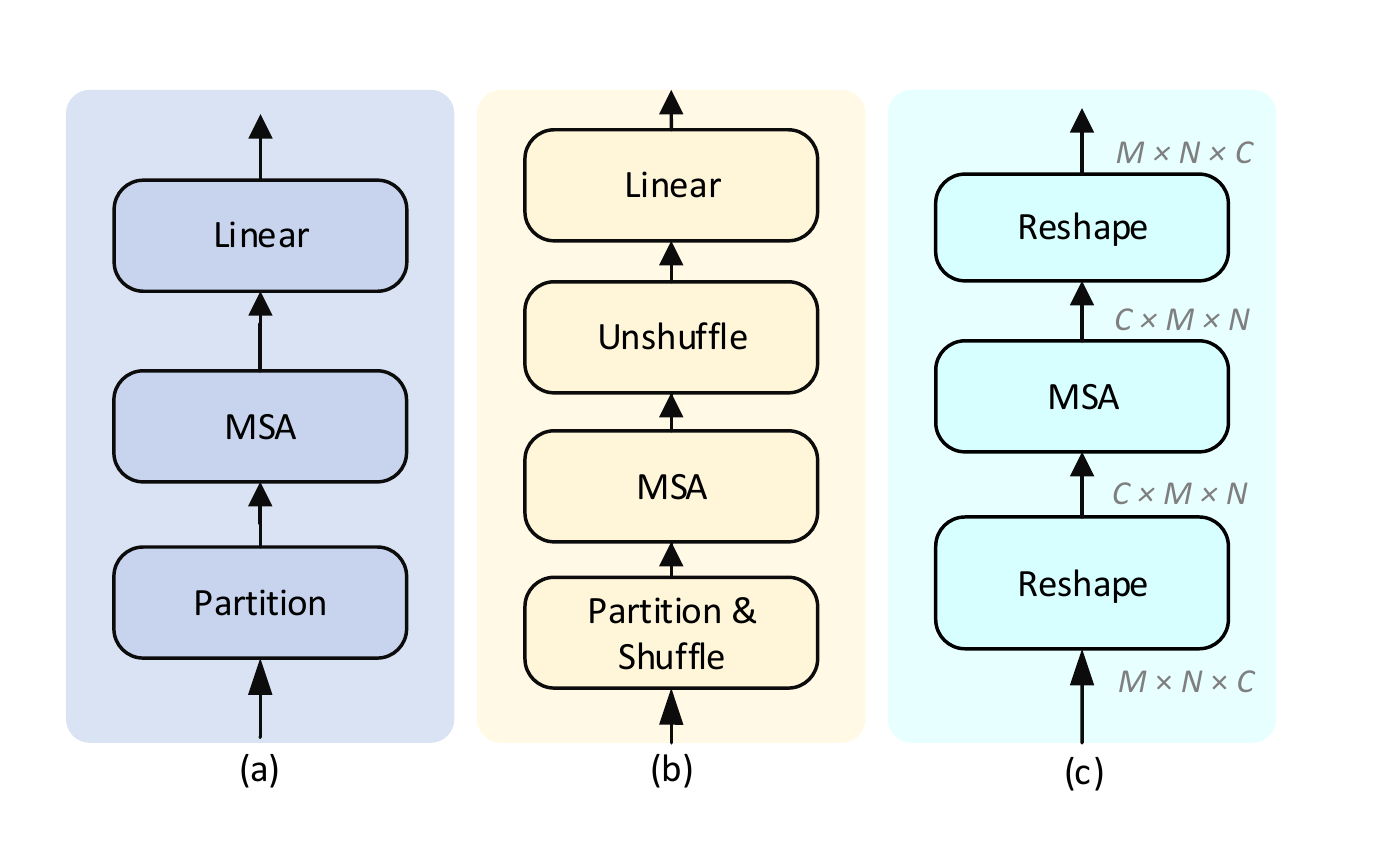}
	\vspace{-12mm}
	\caption{\small Diagram of (a) local attention structure. (b) non-local attention block. (c) spectral attention block, which are the key components of LNSA block.}
	\label{fig:LNSA}
	\vspace{-6mm}
\end{figure}

\noindent \textbf{Local-Nonlocal-Spectral-Attention.} As the basic unit of U-LNSA, LNSA block is customised to model the local, non-local dependencies and spectral correlation of HSI by using a memory efficient group convolution mechanism. 
As depicted in Fig.\ref{fig:U-LNSA} (b), LNSA first feed input into a layer normalization (LN) block and a $conv1\times1$ layer, and then passes it into the parallel local-non-local attention layer and the spectral attention block, respectively. 
Subsequently, the outputs of local attention and non-local attention is concatenated and fused by a $conv 1 \times 1$ layer, the fused outputs is concatenated with the output of spectral attention block and then fused by a $conv 1 \times 1$ layer. Finally, a LN and a feed-forward layer are employed to generate the output of LNSA.

\begin{figure*}[!htbp]
	\centering
	\renewcommand{\h}{0.105}
	\renewcommand{\wa}{0.12}
	\newcommand{\wb}{0.16}
	\renewcommand{\g}{-0.7mm}
	\renewcommand{\tabcolsep}{1.8pt}
	\renewcommand{\arraystretch}{1}
        \resizebox{1.0\linewidth}{!} {
		\begin{tabular}{cc}	
                \Huge
			\renewcommand{\name}{figures/TSA_S1to10/}
			\renewcommand{\h}{0.2}
			\renewcommand{\w}{0.2}
			\begin{tabular}{cc}
				\begin{adjustbox}{valign=t}
					\begin{tabular}{c}%
		         	\includegraphics[trim={0 0 0 0 },clip, width=0.3\textwidth]{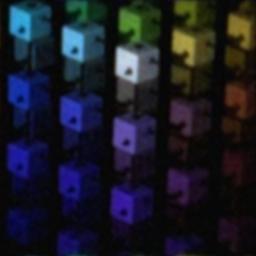}
						\\
						SCENE-01 
					\end{tabular}
				\end{adjustbox}
				\begin{adjustbox}{valign=t}
					\begin{tabular}{cccccccccc}
                            \includegraphics[trim={110 160 90 40  },clip,height=\h \textwidth, width=\w \textwidth]{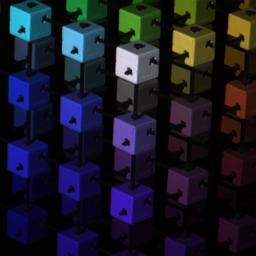} \hspace{\g} &
						\includegraphics[trim={110 160 90 40  },clip,height=\h \textwidth, width=\w \textwidth]{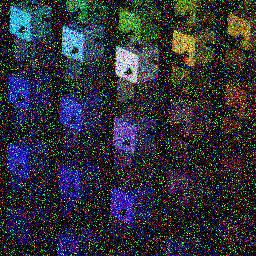} \hspace{\g} &
						\includegraphics[trim={110 160 90 40   },clip,height=\h \textwidth, width=\w \textwidth]{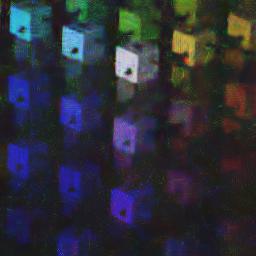} \hspace{\g} &
      					\includegraphics[trim={110 160 90 40  },clip,height=\h \textwidth, width=\w \textwidth]{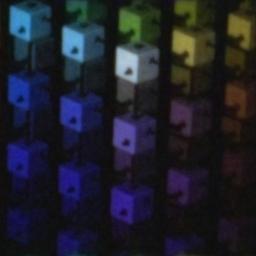} \hspace{\g} &
						\includegraphics[trim={110 160 90 40   },clip,height=\h \textwidth, width=\w \textwidth]{\name 3DTNN_FW_TSA_S1.jpg} \hspace{\g} &
						\includegraphics[trim={110 160 90 40   },clip,height=\h \textwidth, width=\w \textwidth]{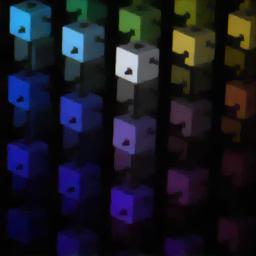} &
						\includegraphics[trim={110 160 90 40   },clip,height=\h \textwidth, width=\w \textwidth]{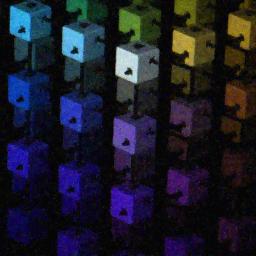} \hspace{\g} &
						\includegraphics[trim={110 160 90 40   },clip,height=\h \textwidth, width=\w \textwidth]{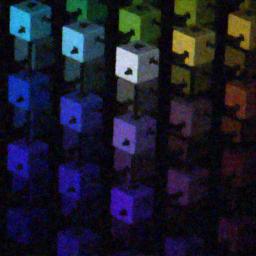} &
						\includegraphics[trim={110 160 90 40   },clip,height=\h \textwidth, width=\w \textwidth]{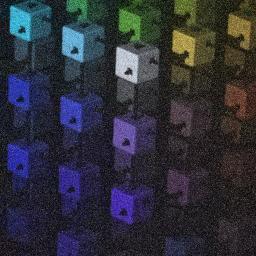} \hspace{\g} &
						\includegraphics[trim={110 160 90 40   },clip,height=\h \textwidth, width=\w \textwidth]{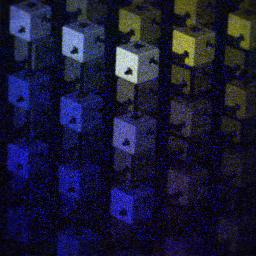} \hspace{\g} 
						\\
						GT \hspace{\g} &
						Noisy Image \hspace{\g} & BM3D~\cite{BM3D} \hspace{\g} & 3DTNN~\cite{3DTNN} \hspace{\g} & 3DTNN-FW~\cite{3DTNN_FW} & E3DTV~\cite{E3DTV} & FGSLR~\cite{FGSLR} & LRTV~\cite{LRTV} & NAILRMA~\cite{NAILRMA} & TLRLSSTV~\cite{TLR_LSSTV}
						\\
                            \includegraphics[trim={110 160 90 40   },clip,height=\h \textwidth, width=\w \textwidth]{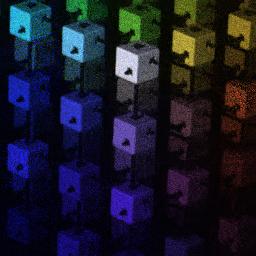} \hspace{\g} &
						\includegraphics[trim={110 160 90 40   },clip,height=\h \textwidth, width=\w \textwidth]{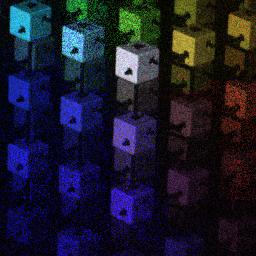} \hspace{\g} &
						\includegraphics[trim={110 160 90 40  },clip,height=\h \textwidth, width=\w \textwidth]{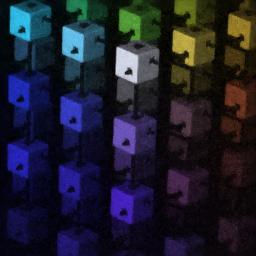} \hspace{\g} &
      					\includegraphics[trim={110 160 90 40   },clip,height=\h \textwidth, width=\w \textwidth]{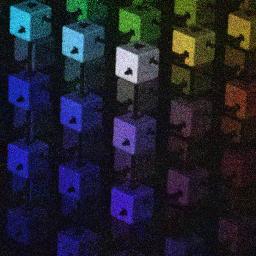} \hspace{\g} &
						\includegraphics[trim={110 160 90 40  },clip,height=\h \textwidth, width=\w \textwidth]{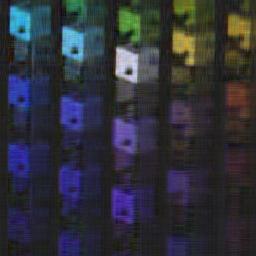} \hspace{\g} &
						\includegraphics[trim={110 160 90 40   },clip,height=\h \textwidth, width=\w \textwidth]{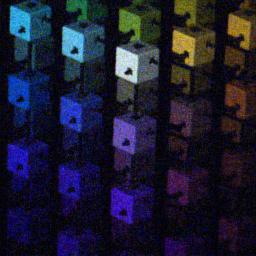}
						\hspace{\g} &		
						\includegraphics[trim={110 160 90 40   },clip,height=\h \textwidth, width=\w \textwidth]{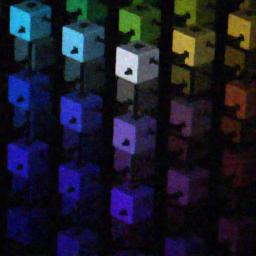} &		
						\includegraphics[trim={110 160 90 40   },clip,height=\h \textwidth, width=\w \textwidth]{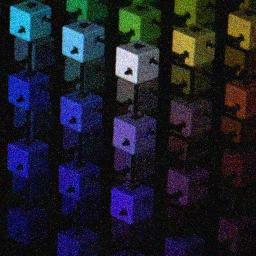} \hspace{\g} &
						\includegraphics[trim={110 160 90 40   },clip,height=\h \textwidth, width=\w \textwidth]{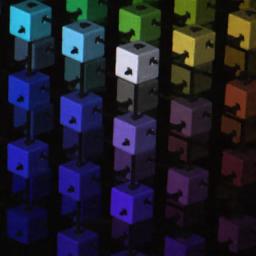}
						\hspace{\g} &
						\includegraphics[trim={110 160 90 40   },clip,height=\h \textwidth, width=\w \textwidth]{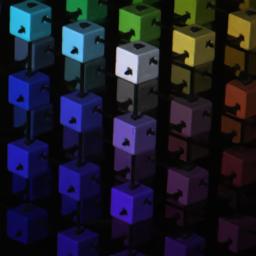} \hspace{\g} 
                            \\
						LLRPnP~\cite{LLRPnP} \hspace{\g} &
						LLRSSTV~\cite{LLRPnP} \hspace{\g} & LLxRGTV~\cite{LLxRGTV} \hspace{\g} & LRMR~\cite{LRMR} \hspace{\g} & LRTA~\cite{LRTA} & LRTDCTV~\cite{LRTDCTV} & LRTDTV~\cite{wang2017hyperspectral} & SSTV~\cite{SSTV} & SERT~\cite{SERT} & \textbf{DNA-Net (Our)}
						\\
					\end{tabular}
				\end{adjustbox}
			\end{tabular}	
		\end{tabular}
	}
        \resizebox{1.0\linewidth}{!} {
		\begin{tabular}{cc}			
			\renewcommand{\name}{figures/TSA_S1to10/}
			\renewcommand{\h}{0.2}
			\renewcommand{\w}{0.2}
			\begin{tabular}{cc}
                    \Huge
				\begin{adjustbox}{valign=t}
					\begin{tabular}{c}%
		         	\includegraphics[trim={0 0 0 0 },clip, width=0.3\textwidth]{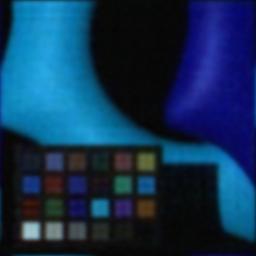}
						\\
						SCENE-03 
					\end{tabular}
				\end{adjustbox}
				\begin{adjustbox}{valign=t}
					\begin{tabular}{cccccccccc}
                            \includegraphics[trim={10 10 190 190  },clip,height=\h \textwidth, width=\w \textwidth]{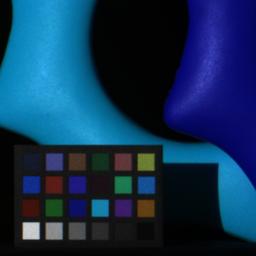} \hspace{\g} &
						\includegraphics[trim={10 10 190 190  },clip,height=\h \textwidth, width=\w \textwidth]{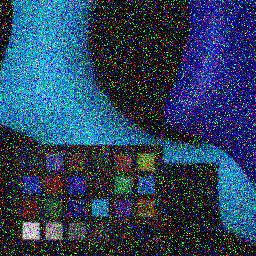} \hspace{\g} &
						\includegraphics[trim={10 10 190 190   },clip,height=\h \textwidth, width=\w \textwidth]{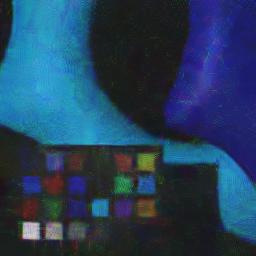} \hspace{\g} &
      					\includegraphics[trim={10 10 190 190  },clip,height=\h \textwidth, width=\w \textwidth]{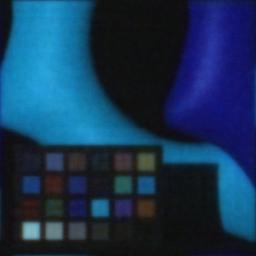} \hspace{\g} &
						\includegraphics[trim={10 10 190 190   },clip,height=\h \textwidth, width=\w \textwidth]{\name 3DTNN_FW_TSA_S3.jpg} \hspace{\g} &
						\includegraphics[trim={10 10 190 190   },clip,height=\h \textwidth, width=\w \textwidth]{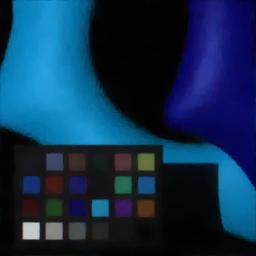} &
						\includegraphics[trim={10 10 190 190   },clip,height=\h \textwidth, width=\w \textwidth]{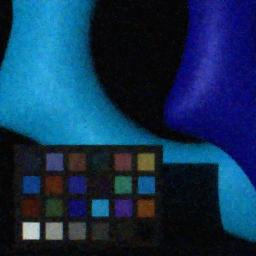} \hspace{\g} &
						\includegraphics[trim={10 10 190 190   },clip,height=\h \textwidth, width=\w \textwidth]{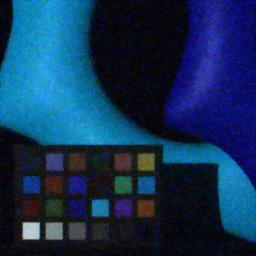} &
						\includegraphics[trim={10 10 190 190   },clip,height=\h \textwidth, width=\w \textwidth]{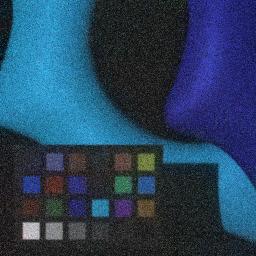} \hspace{\g} &
						\includegraphics[trim={10 10 190 190   },clip,height=\h \textwidth, width=\w \textwidth]{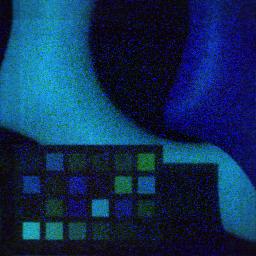} \hspace{\g} 
						\\
						GT \hspace{\g} &
						Noisy Image \hspace{\g} & BM3D~\cite{BM3D} \hspace{\g} & 3DTNN~\cite{3DTNN} \hspace{\g} & 3DTNN-FW~\cite{3DTNN_FW} & E3DTV~\cite{E3DTV} & FGSLR~\cite{FGSLR} & LRTV~\cite{LRTV} & NAILRMA~\cite{NAILRMA} & TLRLSSTV~\cite{TLR_LSSTV}
						\\
                            \includegraphics[trim={10 10 190 190   },clip,height=\h \textwidth, width=\w \textwidth]{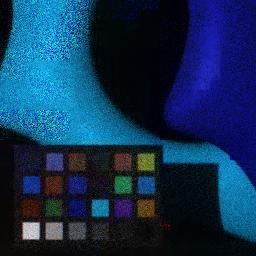} \hspace{\g} &
						\includegraphics[trim={10 10 190 190   },clip,height=\h \textwidth, width=\w \textwidth]{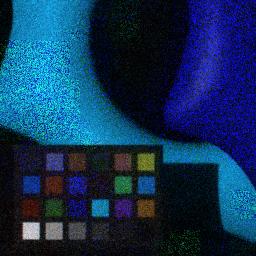} \hspace{\g} &
						\includegraphics[trim={10 10 190 190  },clip,height=\h \textwidth, width=\w \textwidth]{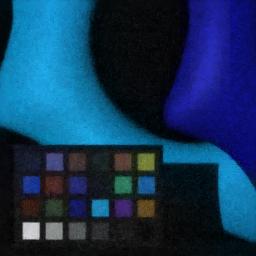} \hspace{\g} &
      					\includegraphics[trim={10 10 190 190   },clip,height=\h \textwidth, width=\w \textwidth]{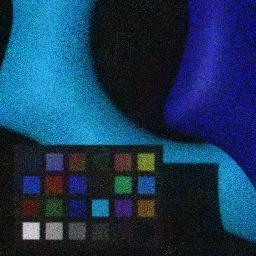} \hspace{\g} &
						\includegraphics[trim={10 10 190 190  },clip,height=\h \textwidth, width=\w \textwidth]{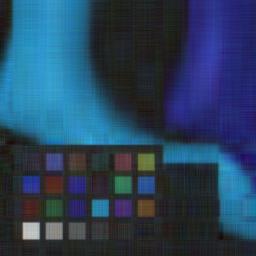} \hspace{\g} &
						\includegraphics[trim={10 10 190 190   },clip,height=\h \textwidth, width=\w \textwidth]{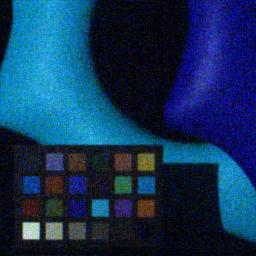}
						\hspace{\g} &		
						\includegraphics[trim={10 10 190 190   },clip,height=\h \textwidth, width=\w \textwidth]{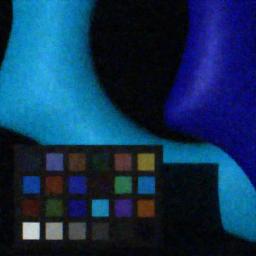} &		
						\includegraphics[trim={10 10 190 190   },clip,height=\h \textwidth, width=\w \textwidth]{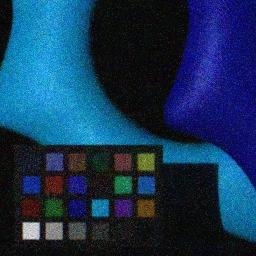} &
						\includegraphics[trim={10 10 190 190   },clip,height=\h \textwidth, width=\w \textwidth]{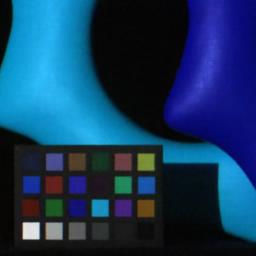}
						\hspace{\g} &
						\includegraphics[trim={10 10 190 190   },clip,height=\h \textwidth, width=\w \textwidth]{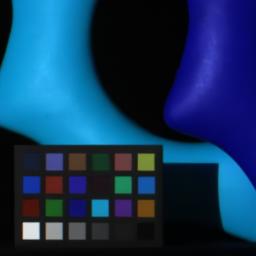} \hspace{\g} 
                            \\
						LLRPnP~\cite{LLRPnP} \hspace{\g} &
						LLRSSTV~\cite{LLRPnP} \hspace{\g} & LLxRGTV~\cite{LLxRGTV} \hspace{\g} & LRMR~\cite{LRMR} \hspace{\g} & LRTA~\cite{LRTA} & LRTDCTV~\cite{LRTDCTV} & LRTDTV~\cite{wang2017hyperspectral} & SSTV~\cite{SSTV} & SERT~\cite{SERT} & \textbf{DNA-Net (Our)}
						\\
					\end{tabular}
				\end{adjustbox}
			\end{tabular}	
		\end{tabular}
	}
	\vspace{-4mm}
	\caption{Visual comparison of \textbf{HSI denoising} methods.} %
	  \vspace{-3mm}
	\label{fig_S1_S3}
\end{figure*}

\noindent \textbf{Local-Non-local Attention Module.} 
The input tokens of local-non-local-attention layer are denoted as $\mathcal{X}_{in}\in \mathbb{R}^{M\times N\times C}$.  Subsequently,  $\mathcal{X}_{in}$ is linearly projected into $query$ $\mathcal{Q}\in \mathbb{R}^{M\times N\times C}$, $key~\mathcal{K} \in \mathbb{R}^{M\times N\times C}$, and $value$ $\mathcal{V}\in \mathbb{R}^{M\times N\times C}$ as
\begin{equation}
	\mathcal{Q}=\mathcal{X}_{in}\mathcal{W^Q},
	~\mathcal{K}=\mathcal{X}_{in}\mathcal{W^K},
	~\mathcal{V}=\mathcal{X}_{in}\mathcal{W^V},
	\label{eq:linear_proj}
 \vspace{-1mm}
\end{equation}
where $\mathcal{W^Q},\mathcal{W^K},\mathcal{W^V}\in \mathbb{R}^{C\times C}$ are learnable parameters and biases are omitted for simplification. 
Then, local contextual information and non-local dependencies are modeled by employing global MSA~\cite{global_msa} and local window and shuffle-based MSA~\cite{liu2021swin,mst}. The implementation of MSAs is parallelized to enhance efficiency. To this end, $\mathcal{Q},\mathcal{K},$ and $\mathcal{V}$ are partitioned into two equal parts along the channel dimension,
\begin{equation}
	\mathcal{Q} = [\mathcal{Q}_1, \mathcal{Q}_{2}], 
	~\mathcal{K} = [\mathcal{K}_1, \mathcal{K}_{2}], 
	~\mathcal{V} = [\mathcal{V}_1, \mathcal{V}_{2}],
	\label{eq:half_split}
\end{equation}
where $\mathcal{Q}_1, \mathcal{K}_1, \mathcal{V}_1\in \mathbb{R}^{M\times N \times \frac{C}{2}}$, $\mathcal{Q}_{2}, \mathcal{K}_{2}, \mathcal{V}_{2}\in \mathbb{R}^{M\times N\times \frac{C}{2}}$. 

Then, $\mathcal{Q}_1,\mathcal{K}_1,\mathcal{V}_1$, $\mathcal{Q}_{2},\mathcal{K}_{2},\mathcal{V}_{2}$ are partitioned into non-overlapping windows of size $p\times p$, and reshaped into $\mathbb{R}^{\frac{MN}{p^2}\times p^2\times \frac{C}{2}}$ firstly. However in non-local branch,
$\mathcal{Q}_{2},\mathcal{K}_{2},\mathcal{V}_{2}$ are transposed from $\mathbb{R}^{\frac{MN}{p^2}\times p^2\times \frac{C}{2}}$ to $\mathbb{R}^{p^2\times \frac{MN}{p^2} \times \frac{C}{2}}$ to shuffle the positions of tokens and establish inter-window dependencies.
Subsequently, 
$\mathcal{Q}_1,\mathcal{K}_1,\mathcal{V}_1$ and $\mathcal{Q}_{2},\mathcal{K}_{2},\mathcal{V}_{2}$ are split along the channel wise into $h$ heads: $\mathcal{Q}_1 = [~\mathcal{Q}_{1}^1, \ldots, \mathcal{Q}_{1}^h~],~\mathcal{K}_1 = [~\mathcal{K}_{1}^1, \ldots, \mathcal{K}_{1}^h~]$, $\mathcal{V}_1 = [~\mathcal{V}_{1}^1, \ldots, \mathcal{V}_{1}^h~]$. The dimension of each head is $d_h = \frac{C}{2h}$. 
 $\mathcal{Q}_{2},\mathcal{K}_{2},\mathcal{V}_{2}$ are split into $h$ heads: $\mathcal{Q}_{2} = [~\mathcal{Q}_{2}^1, \ldots, \mathcal{Q}_{2}^h~],~ \mathcal{K}_{2} = [~\mathcal{K}_{2}^1, \ldots, \mathcal{K}_{2}^h~],$ $\mathcal{V}_{2} =  [~\mathcal{V}_{2}^1, \ldots, \mathcal{V}_{2}^h~]$
The local and nonlocal self-attention $\mathcal{A}_{1}^i$ and $\mathcal{A}_{2}^i$ are calculated inside each head as
\begin{equation}
\begin{aligned}
	\mathcal{A}_{1}^i&=\text{softmax}(\frac{\mathcal{Q}_{1}^i~{\mathcal{K}_{1}^i}\tsp}{\sqrt{d_h}} + \mathcal{P}_{1}^i)~\mathcal{V}_{1}^i, ~~~ i=1, \ldots, h, \\
	\mathcal{A}_{2}^i&=\text{softmax}(\frac{\mathcal{Q}_{2}^i~{\mathcal{K}_{2}^i}\tsp}{\sqrt{d_h}} + \mathcal{P}_{2}^i)~\mathcal{V}_{2}^i, ~~~ i=1, \ldots, h,
	\label{eq:attn_s}
\end{aligned}
\vspace{-2mm}
\end{equation}
where $\mathcal{P}_{1}^i \in \mathbb{R}^{p^2 \times p^2}$, $\mathcal{P}_{2}^i \in \mathbb{R}^{\frac{MN}{p^2} \times \frac{MN}{p^2}}$ are learnable parameters embedding the position information.

\begin{figure*}[t]
	\centering
	\renewcommand{\h}{0.105}
	\renewcommand{\wa}{0.12}
	\newcommand{\wb}{0.16}
	\renewcommand{\g}{-0.7mm}
	\renewcommand{\tabcolsep}{1.8pt}
	\renewcommand{\arraystretch}{1}
                \resizebox{1.0\linewidth}{!} {
		\begin{tabular}{cc}			
			\renewcommand{\name}{figures/TSA_S1to10/}
			\renewcommand{\h}{0.2}
			\renewcommand{\w}{0.2}
			\begin{tabular}{cc}
                \Huge
				\begin{adjustbox}{valign=t}
					\begin{tabular}{c}%
		         	\includegraphics[trim={0 0 0 0 },clip, width=0.3\textwidth]{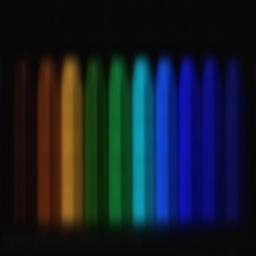}
						\\
						SCENE-09 
					\end{tabular}
				\end{adjustbox}
				\begin{adjustbox}{valign=t}
					\begin{tabular}{cccccccccc}
                            \includegraphics[trim={30 20 170 180  },clip,height=\h \textwidth, width=\w \textwidth]{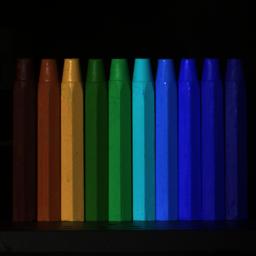} \hspace{\g} &
						\includegraphics[trim={30 20 170 180  },clip,height=\h \textwidth, width=\w \textwidth]{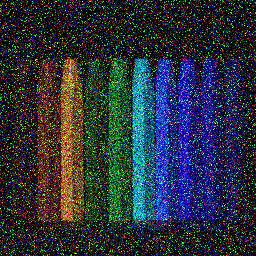} \hspace{\g} &
						\includegraphics[trim={30 20 170 180   },clip,height=\h \textwidth, width=\w \textwidth]{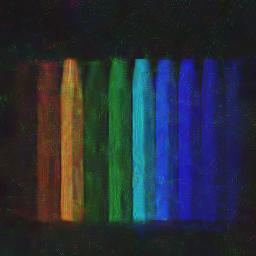} \hspace{\g} &
      					\includegraphics[trim={30 20 170 180  },clip,height=\h \textwidth, width=\w \textwidth]{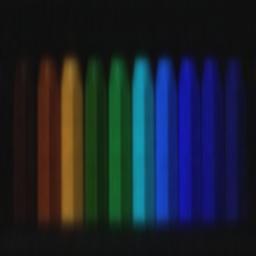} \hspace{\g} &
						\includegraphics[trim={30 20 170 180   },clip,height=\h \textwidth, width=\w \textwidth]{\name 3DTNN_FW_TSA_S9.jpg} \hspace{\g} &
						\includegraphics[trim={30 20 170 180   },clip,height=\h \textwidth, width=\w \textwidth]{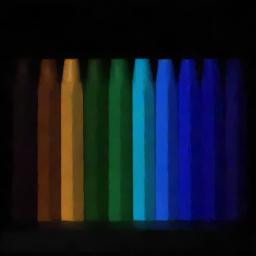} &
						\includegraphics[trim={30 20 170 180   },clip,height=\h \textwidth, width=\w \textwidth]{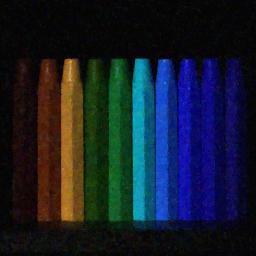} \hspace{\g} &
						\includegraphics[trim={30 20 170 180   },clip,height=\h \textwidth, width=\w \textwidth]{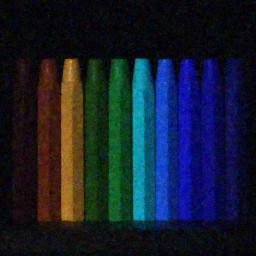} &
						\includegraphics[trim={30 20 170 180   },clip,height=\h \textwidth, width=\w \textwidth]{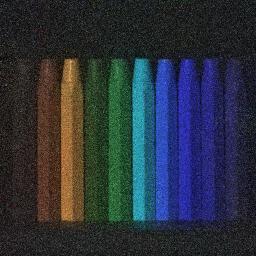} \hspace{\g} &
						\includegraphics[trim={30 20 170 180   },clip,height=\h \textwidth, width=\w \textwidth]{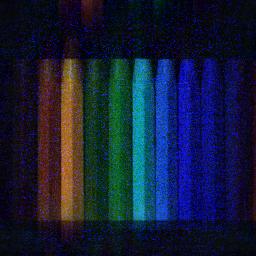} \hspace{\g} 
						\\
						GT \hspace{\g} &
						Noisy Image \hspace{\g} & BM3D~\cite{BM3D} \hspace{\g} & 3DTNN~\cite{3DTNN} \hspace{\g} & 3DTNN-FW~\cite{3DTNN_FW} & E3DTV~\cite{E3DTV} & FGSLR~\cite{FGSLR} & LRTV~\cite{LRTV} & NAILRMA~\cite{NAILRMA} & TLRLSSTV~\cite{TLR_LSSTV}
						\\
                            \includegraphics[trim={30 20 170 180   },clip,height=\h \textwidth, width=\w \textwidth]{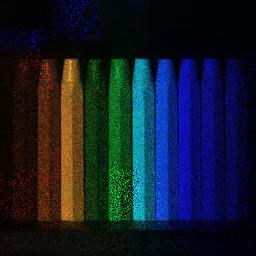} \hspace{\g} &
						\includegraphics[trim={30 20 170 180   },clip,height=\h \textwidth, width=\w \textwidth]{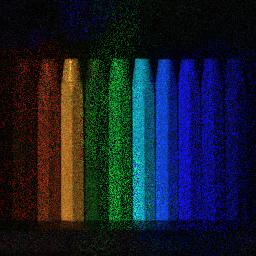} \hspace{\g} &
						\includegraphics[trim={30 20 170 180  },clip,height=\h \textwidth, width=\w \textwidth]{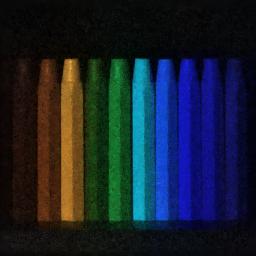} \hspace{\g} &
      					\includegraphics[trim={30 20 170 180   },clip,height=\h \textwidth, width=\w \textwidth]{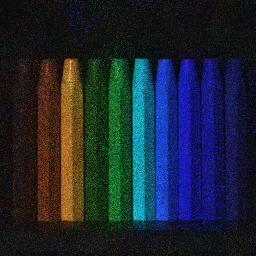} \hspace{\g} &
						\includegraphics[trim={30 20 170 180  },clip,height=\h \textwidth, width=\w \textwidth]{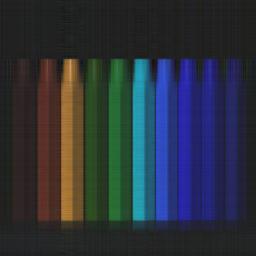} \hspace{\g} &
						\includegraphics[trim={30 20 170 180   },clip,height=\h \textwidth, width=\w \textwidth]{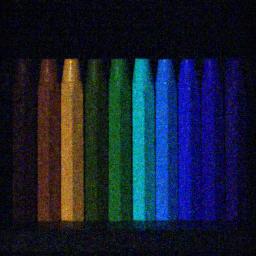}
						\hspace{\g} &		
						\includegraphics[trim={30 20 170 180   },clip,height=\h \textwidth, width=\w \textwidth]{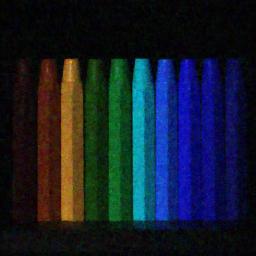} \hspace{\g} &		
						\includegraphics[trim={30 20 170 180   },clip,height=\h \textwidth, width=\w \textwidth]{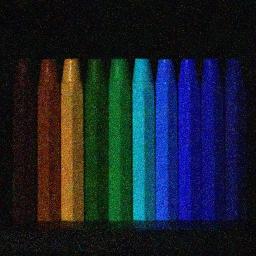} \hspace{\g} &
      					\includegraphics[trim={30 20 170 180   },clip,height=\h \textwidth, width=\w \textwidth]{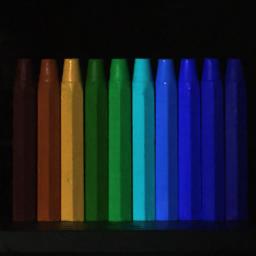}
						\hspace{\g} &
						\includegraphics[trim={30 20 170 180   },clip,height=\h \textwidth, width=\w \textwidth]{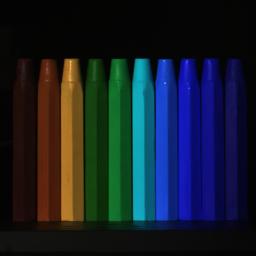} \hspace{\g} 
                            \\
						LLRPnP~\cite{LLRPnP} \hspace{\g} &
						LLRSSTV~\cite{LLRPnP} \hspace{\g} & LLxRGTV~\cite{LLxRGTV} \hspace{\g} & LRMR~\cite{LRMR} \hspace{\g} & LRTA~\cite{LRTA} & LRTDCTV~\cite{LRTDCTV} & LRTDTV~\cite{wang2017hyperspectral} & SSTV~\cite{SSTV} & SERT~\cite{SERT} & \textbf{DNA-Net (Our)}
						\\
					\end{tabular}
				\end{adjustbox}
			\end{tabular}	
		\end{tabular}
	}
        \resizebox{1.0\linewidth}{!} {
		\begin{tabular}{cc}			
			\renewcommand{\name}{figures/TSA_S1to10/}
			\renewcommand{\h}{0.2}
			\renewcommand{\w}{0.2}
			\begin{tabular}{cc}
                    \Huge
				\begin{adjustbox}{valign=t}
					\begin{tabular}{c}%
		         	\includegraphics[trim={0 0 0 0 },clip, width=0.3\textwidth]{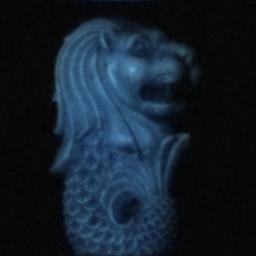}
						\\
						SCENE-10 
					\end{tabular}
				\end{adjustbox}
				\begin{adjustbox}{valign=t}
					\begin{tabular}{cccccccccc}
                            \includegraphics[trim={50 20 150 180  },clip,height=\h \textwidth, width=\w \textwidth]{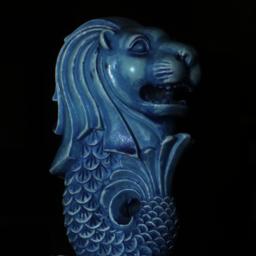} \hspace{\g} &
						\includegraphics[trim={50 20 150 180  },clip,height=\h \textwidth, width=\w \textwidth]{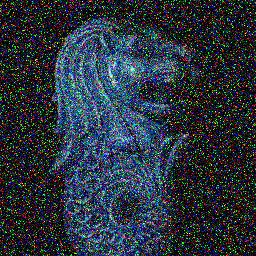} \hspace{\g} &
						\includegraphics[trim={50 20 150 180   },clip,height=\h \textwidth, width=\w \textwidth]{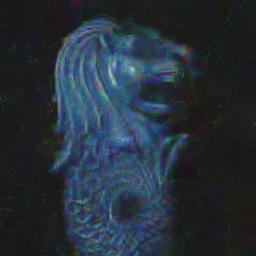} \hspace{\g} &
      					\includegraphics[trim={50 20 150 180  },clip,height=\h \textwidth, width=\w \textwidth]{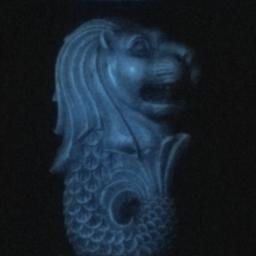} \hspace{\g} &
						\includegraphics[trim={50 20 150 180   },clip,height=\h \textwidth, width=\w \textwidth]{\name 3DTNN_FW_TSA_S10.jpg} \hspace{\g} &
						\includegraphics[trim={50 20 150 180   },clip,height=\h \textwidth, width=\w \textwidth]{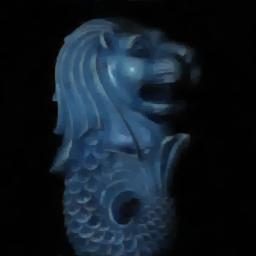} &
						\includegraphics[trim={50 20 150 180   },clip,height=\h \textwidth, width=\w \textwidth]{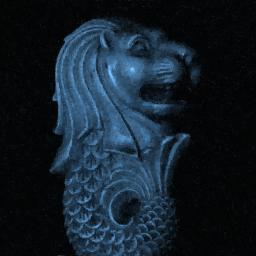} \hspace{\g} &
						\includegraphics[trim={50 20 150 180   },clip,height=\h \textwidth, width=\w \textwidth]{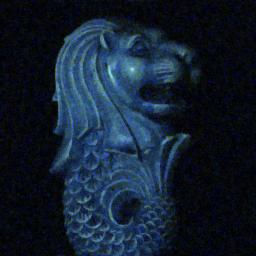} &
						\includegraphics[trim={50 20 150 180   },clip,height=\h \textwidth, width=\w \textwidth]{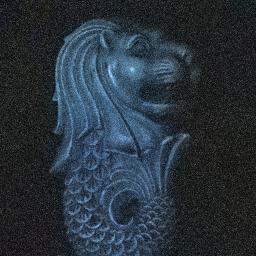} \hspace{\g} &
						\includegraphics[trim={50 20 150 180   },clip,height=\h \textwidth, width=\w \textwidth]{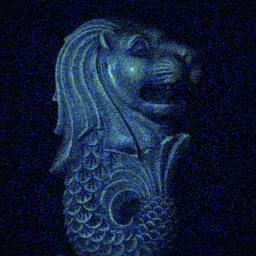} \hspace{\g} 
						\\
						GT \hspace{\g} &
						Noisy Image \hspace{\g} & BM3D~\cite{BM3D} \hspace{\g} & 3DTNN~\cite{3DTNN} \hspace{\g} & 3DTNN-FW~\cite{3DTNN_FW} & E3DTV~\cite{E3DTV} & FGSLR~\cite{FGSLR} & LRTV~\cite{LRTV} & NAILRMA~\cite{NAILRMA} & TLRLSSTV~\cite{TLR_LSSTV}
						\\
                            \includegraphics[trim={50 20 150 180   },clip,height=\h \textwidth, width=\w \textwidth]{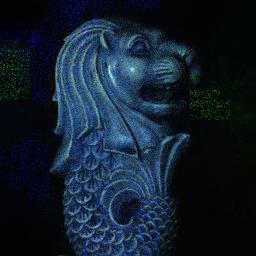} \hspace{\g} &
						\includegraphics[trim={50 20 150 180   },clip,height=\h \textwidth, width=\w \textwidth]{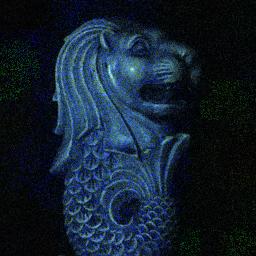} \hspace{\g} &
						\includegraphics[trim={50 20 150 180  },clip,height=\h \textwidth, width=\w \textwidth]{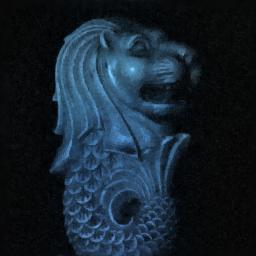} \hspace{\g} &
      					\includegraphics[trim={50 20 150 180   },clip,height=\h \textwidth, width=\w \textwidth]{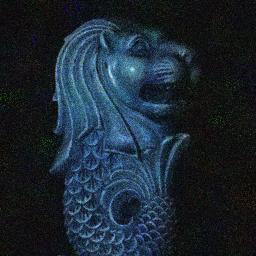} \hspace{\g} &
						\includegraphics[trim={50 20 150 180  },clip,height=\h \textwidth, width=\w \textwidth]{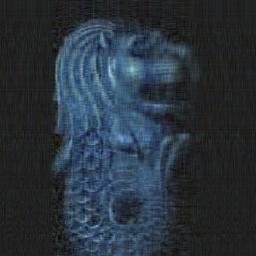} \hspace{\g} &
						\includegraphics[trim={50 20 150 180   },clip,height=\h \textwidth, width=\w \textwidth]{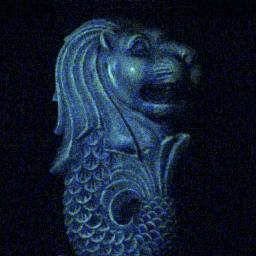}
						\hspace{\g} &		
						\includegraphics[trim={50 20 150 180   },clip,height=\h \textwidth, width=\w \textwidth]{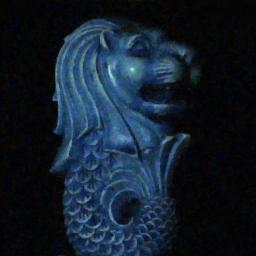} &		
						\includegraphics[trim={50 20 150 180   },clip,height=\h \textwidth, width=\w \textwidth]{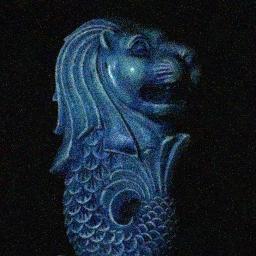} &
      					\includegraphics[trim={50 20 150 180   },clip,height=\h \textwidth, width=\w \textwidth]{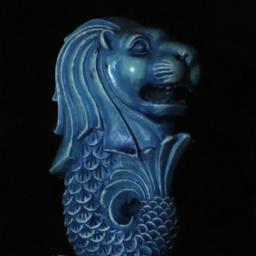}
						\hspace{\g} &
						\includegraphics[trim={50 20 150 180   },clip,height=\h \textwidth, width=\w \textwidth]{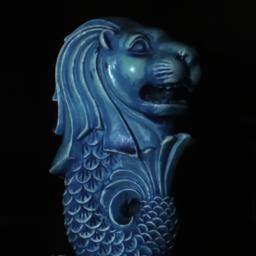} \hspace{\g} 
                            \\
						LLRPnP~\cite{LLRPnP} \hspace{\g} &
						LLRSSTV~\cite{LLRPnP} \hspace{\g} & LLxRGTV~\cite{LLxRGTV} \hspace{\g} & LRMR~\cite{LRMR} \hspace{\g} & LRTA~\cite{LRTA} & LRTDCTV~\cite{LRTDCTV} & LRTDTV~\cite{wang2017hyperspectral} & SSTV~\cite{SSTV} & SERT~\cite{SERT} & \textbf{DNA-Net (Our)}
						\\
					\end{tabular}
				\end{adjustbox}
			\end{tabular}	
		\end{tabular}
	}
	\vspace{-4mm}
	\caption{Visual comparison of \textbf{HSI denoising} methods.} %
	  \vspace{-3mm}
	\label{fig_S9_S10}
\end{figure*}

\vspace{2mm}
\noindent \textbf{Spectral Prior Modeling Module}.
Meanwhile, a Global Spectral Self-Attention module (SM) is utilized in conjunction with the local-non-local branches to effectively capture the essential spectral correlations of HSI. Given an input $\mathcal{X}_{in} \in \mathbb{R}^{H \times W \times C}$, by using SM block, the input is firstly transposed and reshaped into $\mathcal{X}^T \in \mathbb{R}^{C \times H W}$. Subsequently, $X^T$ is linearly projected to $\mathcal{Q}^{sm} \in$ $\mathbb{R}^{C \times H W}, \mathcal{K}^{sm} \in \mathbb{R}^{C \times H W}$, $\mathcal{V}^{sm} \in \mathbb{R}^{C \times H W}$ as:
$
\mathcal{Q}^{sm}=\mathcal{W}_q^{sm} \mathcal{X}^T, \mathcal{K}^{sm}=\mathcal{W}_k^{sm} \mathcal{X}^T$, $\mathcal{V}^{sm}=\mathcal{W}_v^{sm} \mathcal{X}^T,
$
where $\mathcal{W}_q, \mathcal{W}_k, \mathcal{W}_v \in \mathbb{R}^{C \times C}$.
Then, $\mathbf{Q}^{sm}, \mathbf{K}^{sm}$, and $\mathbf{V}^{sm}$ are split into $N$ heads. The spectral attention of each head is computed as 
\begin{equation}
\vspace{-1.5mm}
    \mathcal{A}_j^{sm}= \mathcal{V}_j^{sm} (\operatorname{Softmax}(\mathcal{K}_j^{sm T} \mathcal{Q}_j^{sm} / \sqrt{d})).
\end{equation}
Subsequently, $\mathcal{A}_j^{sm}$ are concatenated in spectral dimension and projected to generate the output $\mathcal{X}_{sm}$.

Finally, $\mathcal{A}_{2}^i \in \mathbb{R}^{p^2\times \frac{MN}{p^2} \times d_h}$ is unshuffled by being transposed to shape $\mathbb{R}^{\frac{MN}{p^2}\times p^2 \times d_h}$, and the final result of LNSA is produced by aggregating the outputs of local-non-local~attention in Eq.~\eqref{eq:attn_s} and spectral prior modeling block,
\vspace{-2mm}
\begin{equation}
	\operatorname{Conv}1 \times 1(\operatorname{Concat}(\sum\nolimits_{i=1}^{h} \mathcal{A}_{1}^i \mathcal{W}_{1}^i+
	\sum\nolimits_{i=1}^{h} \mathcal{A}_{2}^i \mathcal{W}_{2}^i), \mathcal{X}_{sm}),
	\label{eq:head_merge}
	\vspace{-1mm}
\end{equation}
where $\mathcal{W}^i \in \mathbb{R}^{d_h \times C}$ are learnable parameters.

\begin{table*}[!htp]
        \centering
        \caption{Comparisons under noise \emph{Case}-1 and \emph{Case}-2. PSNR, SSIM, FSIM, ERGAS and running time are reported.}
        \vspace{-4mm}
        \setlength{\tabcolsep}{3pt}
	\subfloat[\small Comparisons on S10 under noise \emph{Case}-1.]{ 
		\scalebox{0.9}{
\begin{tabular}{|l|cccc|c|}
\bottomrule[0.1em]
\rowcolor{lightgray}
\textbf{Method} & \textbf{PSNR} $\uparrow $& \textbf{SSIM} $\uparrow$ & \textbf{FSIM} $\uparrow$ & \textbf{ERGAS} $\downarrow$ & \textbf{Time (s)} \\
\toprule[0.1em]
\bottomrule[0.1em]
Noisy & 16.094 & 0.110 & 0.426 & 807.880 & - \\
BWBM3D~\cite{BM3D} & 23.037 & 0.348 & 0.825 & 367.630 & 0.527 \\
LRTA~\cite{LRTA} & 21.302 & 0.267 & 0.691 & 438.080 & 3.205 \\
LRTV~\cite{LRTV} & 28.351 & 0.535 & 0.810 & 205.025 & 9.442 \\
NAILRMA~\cite{NAILRMA} & 23.215 & 0.407 & 0.822 & 356.335 & 8.632 \\
LRMR~\cite{LRMR} & 24.416 & 0.387 & 0.785 & 307.134 & 59.763 \\
NonLRMA~\cite{NonLRMA} & 27.897 & 0.725 & 0.883 & 204.716 & 16.043 \\
LRTDTV~\cite{wang2017hyperspectral} & 30.147 & 0.618 & 0.861 & 159.869 & 43.858 \\
LLRSSTV~\cite{LLRSSTV} & 27.164 & 0.492 & 0.800 & 240.777 & 38.933 \\
SSTV~\cite{SSTV} & 27.052 & 0.456 & 0.832 & 230.285 & 65.186 \\
TLR-LSSTV~\cite{TLR_LSSTV} & 27.981 & 0.539 & 0.821 & 218.898 & 100.600 \\
LLRPnP~\cite{LLRPnP} & 28.735 & 0.608 & 0.897 & 193.391 & 0.530 \\
LLxRGTV~\cite{LLxRGTV} & 28.915 & 0.572 & 0.881 & 184.659 & 39.750 \\
3DTNN~\cite{3DTNN} & 25.326 & 0.436 & 0.848 & 278.152 & 17.160 \\
3DTNN-FW~\cite{3DTNN_FW} & 27.754 & 0.538 & 0.850 & 209.240 & 22.227 \\
LRTDCTV~\cite{LRTDCTV} & 29.289 & 0.593 & 0.822 & 195.385 & 48.965 \\
E3DTV~\cite{E3DTV} & 28.996 & 0.854 & 0.878 & 189.818 & 9.288 \\
FGSLR~\cite{FGSLR}& 30.333 & 0.668 & 0.901 & 157.146 & 192.628 \\
SERT [cvpr 2023]\cite{SERT} & 41.448 & 0.969 & 0.986 & 44.423 & 0.306 \\
\rowcolor{rouse}
DNA-Net-L (Ours)&  41.614 &  0.986 &  0.989 &  43.452 & \bf 0.082 \\
\rowcolor{rouse}
\bf
DNA-Net-B (Ours)& \bf 42.332 & \bf 0.988 & \bf 0.990 & \bf 40.445 & 0.205 \\
\toprule[0.1em]
\end{tabular}}}
\hspace{4.5mm}\vspace{-2mm}
\subfloat[\small Comparisons on S3 under noise \emph{Case}-2.]{ 
\scalebox{0.9}{
\begin{tabular}{|l|cccc|c|}
\bottomrule[0.1em]
\rowcolor{lightgray}
\textbf{Method} & \textbf{PSNR} $\uparrow $& \textbf{SSIM} $\uparrow$ & \textbf{FSIM} $\uparrow$ & \textbf{ERGAS} $\downarrow$ & \textbf{Time (s)} \\
\toprule[0.1em]
\bottomrule[0.1em]
Nosiy & 13.515 & 0.056 & 0.292 & 982.660 & 0.000 \\
BWBM3D~\cite{BM3D} & 23.754 & 0.609 & 0.889 & 318.304 & 0.509 \\
LRTA~\cite{LRTA} & 22.581 & 0.564 & 0.864 & 363.086 & 0.431 \\
LRTV~\cite{LRTV} & 28.978 & 0.588 & 0.769 & 162.753 & 10.316 \\
NAILRMA~\cite{NAILRMA} & 21.765 & 0.359 & 0.710 & 384.629 & 9.010 \\
LRMR~\cite{LRMR} & 24.616 & 0.396 & 0.716 & 257.022 & 63.121 \\
NonLRMA~\cite{NonLRMA} & 20.626 & 0.429 & 0.774 & 447.370 & 17.475 \\
LRTDTV~\cite{wang2017hyperspectral} & 31.199 & 0.682 & 0.828 & 119.344 & 47.167 \\
LLRSSTV~\cite{LLRSSTV} & 25.761 & 0.534 & 0.733 & 230.619 & 41.575 \\ 
SSTV~\cite{SSTV} & 27.014 & 0.433 & 0.727 & 209.430 & 64.225 \\ 
TLRLSSTV~\cite{TLR_LSSTV} & 22.900 & 0.503 & 0.755 & 345.543 & 86.180 \\ 
LLRPnP~\cite{LLRPnP} & 27.206 & 0.634 & 0.807 & 189.142 & 0.583 \\ 
LLxRGTV~\cite{LLxRGTV} & 30.657 & 0.732 & 0.886 & 128.963 & 46.797 \\ 
3DTNN~\cite{3DTNN} & 26.138 & 0.703 & 0.898 & 225.699 & 22.766 \\ 
3DTNN-FW~\cite{3DTNN_FW} & 29.123 & 0.810 & 0.895 & 151.480 & 25.897 \\ 
LRTDCTV~\cite{LRTDCTV} & 27.382 & 0.474 & 0.697 & 192.751 & 65.079 \\ 
E3DTV~\cite{E3DTV} & 32.634 & 0.907 & 0.942 & 168.679 & 11.667 \\ 
FGSLR~\cite{FGSLR} & 31.617 & 0.727 & 0.860 & 115.760 & 285.600 \\
SERT [cvpr 2023]\cite{SERT} & 43.126 & 0.971 & 0.978 & 35.227 & 0.302 \\
\rowcolor{rouse}
DNA-Net-L (Ours)&  43.888 &  0.980 &  0.979 &  31.632 & \bf 0.091 \\
\rowcolor{rouse}
\bf
DNA-Net-B (Ours)& \bf 44.658 & \bf 0.981 & \bf 0.979 & \bf 30.165 & 0.214 \\
\toprule[0.1em]
\end{tabular}}}\hspace{4mm}
	\label{table:case1-2}\vspace{-6mm}
\end{table*}

\vspace{-0.5mm}
\section{Experiment} \label{sec:exp}
\vspace{-0.5mm}
\subsection{Experiment Setup}
\vspace{-0.5mm}

We employed HSI consisting of 28 wavelengths ranging from 450nm to 650nm, obtained through spectral interpolation. The performance of the proposed method was evaluated through both simulation and real experiments.
Simulation experiments were conducted using two datasets, namely CAVE~\cite{cave} and KAIST~\cite{kaist}. The CAVE dataset comprised 32 HSIs with spatial dimensions of 512$\times$512, while the KAIST dataset contained 30 HSIs with spatial dimensions of 2704$\times$3376. The CAVE dataset was utilized as the training set, whereas 10 scenes from the KAIST dataset were selected for testing purposes.
Real experiments were conducted using Urban which had real noise. 
Pytorch was used to implement the proposed DNA-Net, which was trained with the Adam optimizer \cite{adam} and $\beta_1$ = 0.9 and $\beta_2$ = 0.999, for 110 epochs on a 2$\times$GTX 1080Ti 11GB GPU using the Cosine Annealing scheme \cite{cosine}. 
The training objective was to minimize the Root Mean Square Error (RMSE) between the denoised HSIs and the corresponding ground-truth HSIs.

\vspace{-2mm}
\subsection{Quantitative Comparisons with State-of-the-Art Methods}

We compare the results of DNA-Net-B with 5 iterations, its light version DNA-Net-L with 2 iterations, and 18 SOTA methods including BM3D~\cite{BM3D}, LRTA~\cite{LRTA}, and LRTV~\cite{LRTV}), NAILRMA~\cite{NAILRMA}, LRMR~\cite{LRMR}, NonLRMA~\cite{NonLRMA}, LRTDTV~\cite{wang2017hyperspectral}, LLRSSTV~\cite{LLRSSTV}, SSTV~\cite{SSTV}, TLR-LSSTV~\cite{TLR_LSSTV}, and LLRPnP~\cite{LLRPnP}, LLxRGTV~\cite{LLxRGTV}, 3DTNN~\cite{3DTNN}, 3DTNN-FW~\cite{3DTNN_FW}, LRTDCTV~\cite{LRTDCTV}, E3DTV~\cite{E3DTV}, FGSLR~\cite{FGSLR}, and SERT~\cite{SERT} on 10 simulation scenes and real noisy HSI-Urban.

\noindent \textbf{Simulation HSI Denoising.} 
\emph{Noise cases}.
\emph{Case}-1: Gaussian noise $\mathcal{N}(0, 0.2)$, \emph{Case}-2: Gaussian noise $\mathcal{N}(0, 0.2)$ + sparse (impulse) noise with $p=0.05$, \emph{Case}-3: Gaussian noise $\mathcal{N}(0, 0.2)$ + sparse noise with $p=0.1$. 
\emph{Case}-4: Gaussian noise $\mathcal{N}(0, 0.2)$ + sparse noise with $p=0.15$.
\emph{Evaluation metrics}. We employed the widely-used quantitative picture
quality indices (PQIs): PSNR, SSIM, FSIM~\cite{FSIM}, ERGAS~\cite{ERGAS} metrics to evaluate the denoising performance quantitatively.

Tables \ref{table:case1-2} and \ref{table:case3-4} depict the outcomes of experiments on scenes 1 to 10 of the KAIST dataset, aimed at removing mixed Gaussian and sparse noise. Our proposed DNA-Net model was compared against several SOTA denoising approaches. The DNA-Net-B model attained PSNR values of 42.33 dB, 44.66 dB, 44.17 dB, and 43.78 dB, surpassing all other methods by an average of 1.49 dB. Moreover, with respect to SSIM, DNA-Net outperformed the recently proposed state-of-the-art SERT method~\cite{SERT} by a minimum of 0.1.

Visual analysis of the results is presented in Fig. \ref{fig_S1_S3} and \ref{fig_S9_S10}, which demonstrate the efficacy of DNA-Net in reducing mixed noise while preserving the details of HSI. To evaluate whether the enhanced performance was obtained at the expense of increased computational costs, we compared the PQIs results against computational cost, which are shown in Tables \ref{table:case1-2} and \ref{table:case3-4}. The DNA-Net-L model delivered superior performance compared to most models, with the lowest computational cost, highlighting its efficiency and effectiveness.

\begin{figure*}[!htp]
	\centering
        \Large
	\renewcommand{\h}{0.105}
	\renewcommand{\wa}{0.12}
	\newcommand{\wb}{0.16}
	\renewcommand{\g}{-0.7mm}
	\renewcommand{\tabcolsep}{1.8pt}
	\renewcommand{\arraystretch}{1}
        \resizebox{1.0\linewidth}{!} {
		\begin{tabular}{cc}			
			\renewcommand{\name}{figures/urban_real/}
			\renewcommand{\h}{0.2}
			\renewcommand{\w}{0.2}
			\begin{tabular}{cc}
                    \Huge
				\begin{adjustbox}{valign=t}
					\begin{tabular}{c}%
		         	\includegraphics[trim={0 0 0 0 },clip, width=0.3\textwidth]{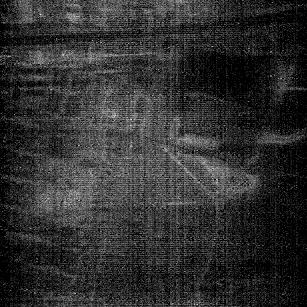}
						\\
						Urban-108 
					\end{tabular}
				\end{adjustbox}
				\begin{adjustbox}{valign=t}
					\begin{tabular}{ccccccccc}
						\includegraphics[trim={150 50 50 150  },clip,height=\h \textwidth, width=\w \textwidth]{\name Nosiy_urban_real_b108.jpg} \hspace{\g} &
      					\includegraphics[trim={150 50 50 150  },clip,height=\h \textwidth, width=\w \textwidth]{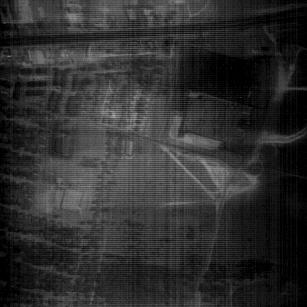} \hspace{\g} &
						\includegraphics[trim={150 50 50 150   },clip,height=\h \textwidth, width=\w \textwidth]{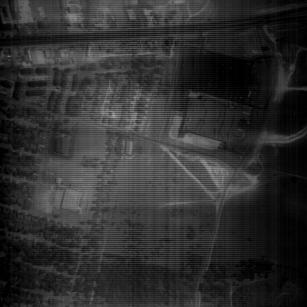} \hspace{\g} &
						\includegraphics[trim={150 50 50 150   },clip,height=\h \textwidth, width=\w \textwidth]{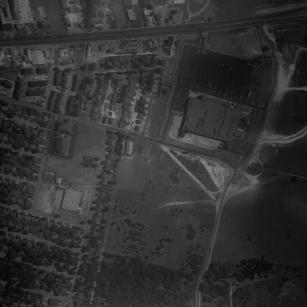} &
						\includegraphics[trim={150 50 50 150   },clip,height=\h \textwidth, width=\w \textwidth]{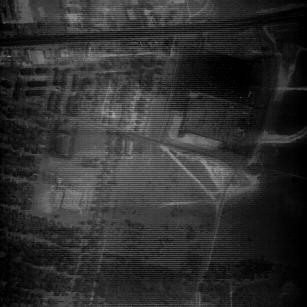} \hspace{\g} &
						\includegraphics[trim={150 50 50 150   },clip,height=\h \textwidth, width=\w \textwidth]{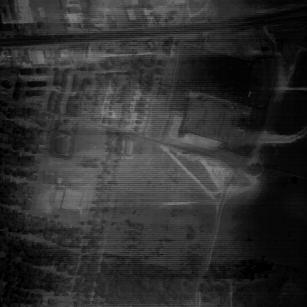} &
						\includegraphics[trim={150 50 50 150   },clip,height=\h \textwidth, width=\w \textwidth]{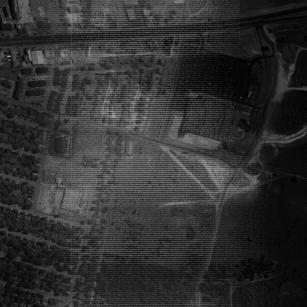} \hspace{\g} &
						\includegraphics[trim={150 50 50 150   },clip,height=\h \textwidth, width=\w \textwidth]{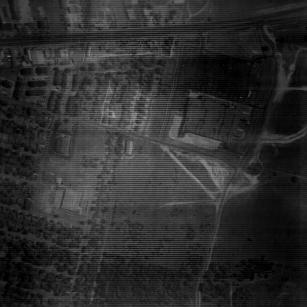} \hspace{\g} &
                         \includegraphics[trim={150 50 50 150   },clip,height=\h \textwidth, width=\w \textwidth]{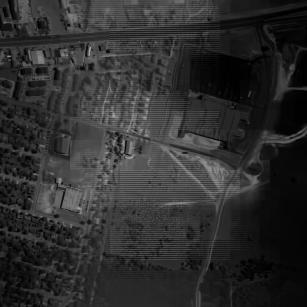} \hspace{\g} 
						\\
						Noisy \hspace{\g} & 3DTNN~\cite{3DTNN} \hspace{\g} & 3DTNN-FW~\cite{3DTNN_FW} & E3DTV~\cite{E3DTV} & FGSLR~\cite{FGSLR} & LRTV~\cite{LRTV} & NAILRMA~\cite{NAILRMA} & TLRLSSTV~\cite{TLR_LSSTV}  & LLRPnP~\cite{LLRPnP} \hspace{\g} 
						\\
						\includegraphics[trim={150 50 50 150   },clip,height=\h \textwidth, width=\w \textwidth]{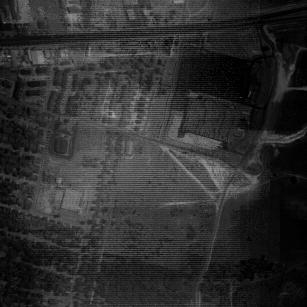} \hspace{\g} &
						\includegraphics[trim={150 50 50 150  },clip,height=\h \textwidth, width=\w \textwidth]{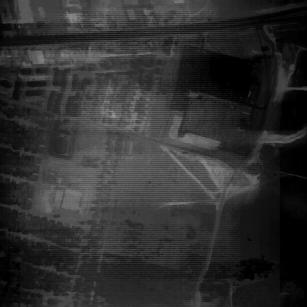} \hspace{\g} &
      					\includegraphics[trim={150 50 50 150   },clip,height=\h \textwidth, width=\w \textwidth]{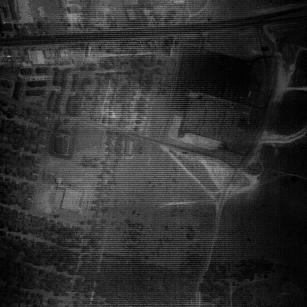} \hspace{\g} &
						\includegraphics[trim={150 50 50 150  },clip,height=\h \textwidth, width=\w \textwidth]{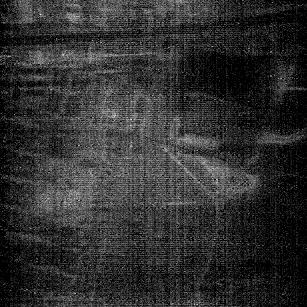} \hspace{\g} &
						\includegraphics[trim={150 50 50 150   },clip,height=\h \textwidth, width=\w \textwidth]{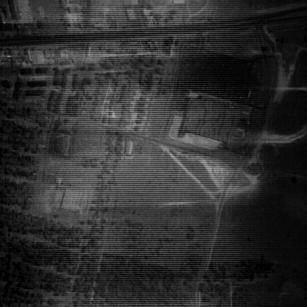}
						\hspace{\g} &		
						\includegraphics[trim={150 50 50 150   },clip,height=\h \textwidth, width=\w \textwidth]{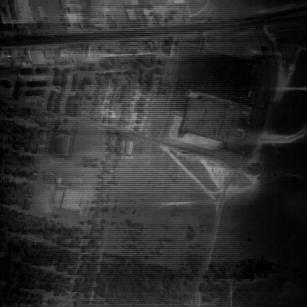} \hspace{\g} &		
						\includegraphics[trim={150 50 50 150   },clip,height=\h \textwidth, width=\w \textwidth]{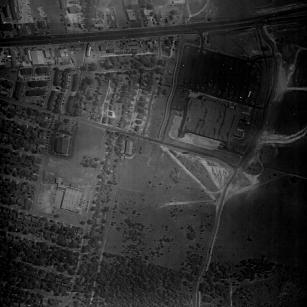} \hspace{\g} &
      					\includegraphics[trim={150 0 0 150   },clip,height=\h \textwidth, width=\w \textwidth]{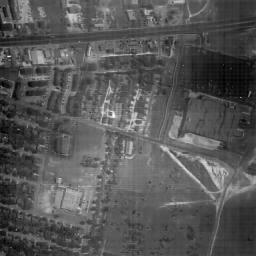}
						\hspace{\g} &
						\includegraphics[trim={150 0 0 150   },clip,height=\h \textwidth, width=\w \textwidth]{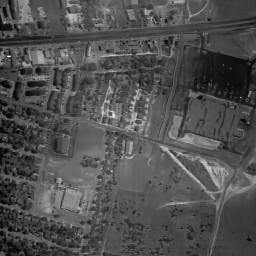} \hspace{\g} 
                            \\
						LLRSSTV~\cite{LLRSSTV} \hspace{\g} & LLxRGTV~\cite{LLxRGTV} \hspace{\g} & LRMR~\cite{LRMR} \hspace{\g} & LRTA~\cite{LRTA} & LRTDCTV~\cite{LRTDCTV} & LRTDTV~\cite{wang2017hyperspectral} & SSTV~\cite{SSTV} & SERT~\cite{SERT}  & \textbf{DNA-Net (Our)}
						\\
					\end{tabular}
				\end{adjustbox}
			\end{tabular}	
		\end{tabular}
	}
	\vspace{-3mm}
	\caption{Visual comparison of \textbf{HSI denoising} methods on real noisy HSI Urban.} %
	  \vspace{-3mm}
	\label{fig_urban_real}
\end{figure*}

\begin{table*}
        \centering
        \caption{\small Ablation studies on \emph{Scene} 10 of simulation datasets~\cite{kaist}. PSNR, SSIM, FSIM, ERGAS are reported.}
        \vspace{-3mm}
		\scalebox{1.08}{
			\begin{tabular}{c c c  c c c c c }
				\bottomrule
				\rowcolor{color3} DNA-Net & Local Attention & Non-local Attention  & Spectral Attention & PSNR $\uparrow $& SSIM $\uparrow$ & FSIM $\uparrow$ & ERGAS $\downarrow$ \\
				\midrule
    			\checkmark & \checkmark & &  &40.300 &0.972 &0.981 &50.089 \\
				\checkmark & \checkmark & & \checkmark &41.247 &0.978 &0.986 &45.116 \\
				\checkmark  &\checkmark & \checkmark &&40.563 &0.973 &0.985 &49.236 \\
				\checkmark  &\checkmark & \checkmark &\checkmark &\bf 41.489 &\bf 0.986 & \bf 0.988 & \bf 44.106 \\
				\bottomrule
	\end{tabular}}\hspace{2mm}\vspace{-1.5mm}
	\label{tab:ablations}\vspace{-2mm}
\end{table*}

\noindent \textbf{Real HSI Denoising.} 
The HYDICE Urban hyperspectral imagery dataset \footnote{http://www.tec.army.mil/
hypercube/} has an original size of 207 $\times$ 207 $\times$ 210, with 189 bands remaining after excluding the water absorption band. Fig. \ref{fig_urban_real} displays grayscale images of the 108th band in the noisy HSI and the results after applying various restoration methods. Among these methods, the DNA-Net method is particularly effective in removing sparse noise, especially strip noise, and restoring spatial details, resulting in the best visual quality of the restored image.

\begin{table*}[t]
        \centering
        \caption{Comparisons under noise \emph{Case}-3 and \emph{Case}-4. PSNR, SSIM, FSIM, ERGAS and running time are reported.}
        \vspace{-4mm}
        \setlength{\tabcolsep}{3pt}
	\subfloat[\small Comparisons on S1 under noise \emph{Case}-3.]{ 
		\scalebox{0.9}{
\begin{tabular}{|l|cccc|c|}
\bottomrule[0.1em]
\rowcolor{lightgray}
\textbf{Method} & \textbf{PSNR} $\uparrow $& \textbf{SSIM} $\uparrow$ & \textbf{FSIM} $\uparrow$ & \textbf{ERGAS} $\downarrow$ & \textbf{Time (s)} \\
\toprule[0.1em]
\bottomrule[0.1em]
Noisy & 11.830 & 0.045 & 0.331 & 1324.192  & - \\
BWBM3D~\cite{BM3D} & 21.309 & 0.449 & 0.790 & 445.832 &  0.611 \\
LRTA~\cite{LRTA}& 20.310 & 0.481 & 0.827 & 499.671 &  0.479 \\
LRTV~\cite{LRTV} & 28.872 & 0.617 & 0.806 & 225.460  & 9.302 \\
NAILRMA \cite{NAILRMA}& 20.055 & 0.318 & 0.708 & 513.617  & 10.896 \\
LRMR~\cite{LRMR}& 24.204 & 0.400 & 0.745 & 317.549  & 60.537 \\
NonLRMA~\cite{NonLRMA} & 22.746 & 0.388 & 0.773 & 376.698  & 9.768 \\
LRTDTV~\cite{wang2017hyperspectral}& 30.629 & 0.695 & 0.848 & 154.044  & 43.426 \\
LLRSSTV~\cite{LLRSSTV} & 26.010 & 0.510 & 0.754 & 288.215  & 40.266 \\
SSTV~\cite{SSTV}& 26.669 & 0.474 & 0.781 & 239.868  & 63.276 \\
TLR-LSSTV~\cite{TLR_LSSTV} & 25.284 & 0.461 & 0.732 & 407.490  & 85.206 \\
LLRPnP~\cite{LLRPnP} & 27.275 & 0.643 & 0.832 & 243.813  & 0.748 \\
LLxRGTV\cite{LLxRGTV} & 29.296 & 0.705 & 0.890 & 176.948  & 45.746 \\
3DTNN\cite{3DTNN} & 26.739 & 0.671 & 0.863 & 237.061  & 20.755 \\
3DTNN-FW\cite{3DTNN_FW} & 29.730 & 0.775 & 0.869 & 168.599 & 25.228 \\
LRTDCTV~\cite{LRTDCTV} & 28.187 & 0.562 & 0.785 & 231.739 & 59.314 \\
E3DTV~\cite{E3DTV} & 30.820 & 0.852 & 0.876 & 151.999 & 10.955 \\
FGSLR~\cite{FGSLR} & 30.909 & 0.729 & 0.872 & 146.890 & 379.678 \\
SERT [cvpr 2023]\cite{SERT} & 42.350 & 0.969 & 0.984 & 45.665 & 0.293 \\
\rowcolor{rouse}
DNA-Net-L (Ours)&  43.072 &  0.981 &  0.988 &  42.095 & \bf 0.086 \\
\rowcolor{rouse}
\bf
DNA-Net-B (Ours)& \bf 44.168 & \bf 0.983 & \bf 0.989 & \bf 37.242 & 0.212 \\
\toprule[0.1em]
\end{tabular}
}}
\hspace{2mm}\vspace{-1.5mm}
\subfloat[\small Comparisons on S9 under noise \emph{Case}-4.]{ 
		\scalebox{0.9}{
\begin{tabular}{|l|cccc|c|}
\bottomrule[0.1em]
\rowcolor{lightgray}
\textbf{Method} & \textbf{PSNR} $\uparrow $& \textbf{SSIM} $\uparrow$ & \textbf{FSIM} $\uparrow$ & \textbf{ERGAS} $\downarrow$ & \textbf{Time (s)} \\
\toprule[0.1em]
\bottomrule[0.1em]
Noisy & 10.688 & 0.034 & 0.247 & 1493.682  & - \\
BWBM3D~\cite{BM3D} & 19.239 & 0.281 & 0.677 & 558.502 & 0.563 \\
LRTA~\cite{LRTA} & 18.721 & 0.465 & 0.882 & 595.824 & 0.272 \\
LRTV~\cite{LRTV} & 27.957 & 0.535 & 0.729 & 217.996 & 9.498 \\
NAILRMA~\cite{NAILRMA} & 17.972 & 0.233 & 0.613 & 646.518 & 11.025 \\
LRMR~\cite{LRMR} & 22.666 & 0.355 & 0.669 & 372.022 & 60.654 \\
NonLRMA~\cite{NonLRMA} & 21.950 & 0.453 & 0.747 & 453.460 & 13.077 \\
LRTDTV~\cite{wang2017hyperspectral} & 30.086 & 0.637 & 0.788 & 161.100 & 47.716 \\
LLRSSTV~\cite{LLRSSTV} & 23.405 & 0.546 & 0.700 & 361.783 & 40.168 \\
SSTV~\cite{SSTV} & 26.341 & 0.406 & 0.707 & 244.761 & 64.891 \\
TLRLSSTV~\cite{TLR_LSSTV}  & 26.008 & 0.585 & 0.763 & 286.956 & 87.575 \\
LLRPnP~\cite{LLRPnP}  & 25.414 & 0.674 & 0.791 & 287.431 & 0.599 \\
LLxRGTV~\cite{LLxRGTV}  & 29.750 & 0.682 & 0.875 & 163.572 & 48.340 \\
3DTNN~\cite{3DTNN}  & 26.068 & 0.643 & 0.897 & 251.031 & 22.845 \\
3DTNN-FW~\cite{3DTNN_FW}  & 28.953 & 0.759 & 0.882 & 179.162 & 28.159 \\
LRTDCTV~\cite{LRTDCTV}  & 25.592 & 0.398 & 0.644 & 279.517 & 67.074 \\
E3DTV~\cite{E3DTV}  & 30.343 & 0.879 & 0.915 & 189.866 & 11.735 \\
FGSLR~\cite{FGSLR}  & 29.914 & 0.632 & 0.804 & 160.725 & 491.061 \\
SERT [cvpr 2023]\cite{SERT} & 41.987 & 0.969 & 0.977 & 42.999 & 0.318 \\
\rowcolor{rouse}
DNA-Net-L (Ours)&  41.952 &  0.980 &  0.981 &  43.734 & \bf 0.086 \\
\rowcolor{rouse}
\bf
DNA-Net-B (Ours)& \bf 43.780 & \bf 0.983 & \bf 0.984 & \bf 35.252 & 0.210 \\
\toprule[0.1em]
\end{tabular}}}\hspace{4mm}
	\label{table:case3-4}\vspace{-6mm}
\end{table*}

\vspace{-1mm}
\section{Ablation Study}
\textbf{Component Ablation.} Table \ref{tab:ablations} presents the results of various component designs. The first row corresponds to the proposed DNA-Net that solely utilizes local attention in the spatial domain. The 2nd, 3rd, and 4th rows represent DNA-Net models that incorporate additional spectral attention, non-local attention, and spectral plus non-local attention, respectively. The results demonstrate that incorporating spectral and non-local attention blocks to capture spatial-spectral information significantly improves the denoising performance.

\noindent \textbf{Iteration Analysis.} In Tables \ref{table:case1-2} and \ref{table:case3-4}, the performance of DNA-Net with 2 and 5 iterations, denoted as DNA-Net-L and DNA-Net-B, respectively, is presented in the last two rows. The results indicate that DNA-Net-B displays an enhancement in PQIs, albeit at the expense of increased running time. Therefore, selecting the appropriate iteration model for a given application in practical scenarios requires a trade-off between speed and performance.

\vspace{-2mm}
\section{Conclusion} 
This paper addresses three issues that arise in previous model-based or learning-based HSI denoising methods. These issues include a lack of informative parameter estimation for iterative learning, the inability to model noise distributions, and limitations in capturing long-range dependencies due to their primary reliance on CNN-based approaches. To overcome these challenges, we propose a new MAP-based unfolding framework called DNA-Net. The DNA-Net framework is capable of effectively modeling both sparse and Gaussian noise distributions and accurately estimating parameters from a noisy HSI and degradation matrix. These estimated parameters are subsequently utilized to contextualize the effect of noise terms and provide important noise level information for the denoising network during each iteration.
Furthermore, we introduce a new Transformer-based method, U-LNSA, which is capable of jointly extracting spectral correlation, presenting local contents, and modeling non-local dependencies. By integrating U-LNSA into the DNA-Net framework, we present the first Transformer-based unfolding method for HSI denoising. Experimental results clearly indicate that the proposed DNA-Net approach outperforms state-of-the-art methods by a significant margin while requiring much lower memory and computational costs.

\bibliographystyle{ACM-Reference-Format}
\bibliography{reference}


\end{document}